\documentclass{jpp}
\usepackage{url}
\usepackage{graphicx}
\usepackage{subcaption}
\usepackage[utf8]{inputenc}
\usepackage[T1]{fontenc}
\usepackage{amsmath}
\usepackage{breqn}
\usepackage{placeins}
\usepackage{enumitem}
\usepackage{nameref}
\usepackage{hyperref}

\usepackage[table]{xcolor}
\usepackage{makecell}

\newlist{gap_detect}{enumerate}{1}

\makeatletter
\newcommand{\gapitem}[3]{%
    \item[\textbf{#1}]%
    \protected@edef\@currentlabel{#1~#3}%
    \phantomsection\label{gap:#2}%
    \textbf{#3}%
}
\newcommand{\gapref}[1]{\ref{gap:#1}}
\makeatother

\usepackage{bm}
\usepackage{xcolor}

\newcommand{\deltan}{\delta\!n}
\newcommand{\deltam}{\delta\!m}

\newcommand{\secref}[1]{Section~\ref{#1}}
\newcommand{\tabref}[1]{Table~\ref{#1}}
\newcommand{\figref}[1]{Figure~\ref{#1}}
\newcommand{\appref}[1]{Appendix~\ref{#1}}

\shorttitle{The shear Alfv\'{e}n continuum of optimized stellarators}
\shortauthor{A. Hyder, E. Paul, A. Knyazev}

\title{The shear Alfv\'{e}n continuum of optimized stellarators}

\author{Abdullah S. Hyder\aff{1}, Elizabeth J. Paul\aff{1}, Alex R. Knyazev\aff{1}}
\shortauthor{A.S. Hyder, E.J. Paul, A.R. Knyazev}

\affiliation{\aff{1}Department of Applied Physics and Applied Mathematics, Columbia University, New York, NY 10027, USA}

\begin{document}

\maketitle
\thispagestyle{plain}

\begin{abstract}
The shear Alfvén continuum has been studied extensively in tokamaks, but its gap structure in modern optimized stellarators has received less attention, despite its importance for energetic-particle confinement in fusion pilot plants. We compute and compare the continua of five reactor-scale stellarator equilibria: two quasi-axisymmetric (QA), two quasi-helically symmetric (QH), and one quasi-isodynamic (QI). We present selective Fourier mode schemes that provide a sufficient resolution of eigenfrequencies within the frequency range of interest. In all five configurations, the widest gap is the HAE(2,$N_{\text{fp}}$) gap, which is driven by rotating ellipticity. This gap lies in the frequency range of strong resonance with prompt fusion-born alphas in the QA configurations but not in the QH and QI configurations, suggesting that QH and QI configurations may be easier to optimize for alpha confinement. Near-axis theory predicts the HAE(2,$N_{\text{fp}}$) and EAE gap widths well but overestimates the MAE gap widths and the gap widths of the QI configuration. These discrepancies likely arise from interactions between gaps, which depend on the rotational transform profile and are not captured by near-axis theory.
\end{abstract}

\section{Introduction}
\label{sec:intro}
The shear Alfv\'{e}n wave is a transverse low-frequency magnetohydrodynamic (MHD) wave \citep{Liu_Chen_1995}. Energetic particles (EPs), such as fusion-born alpha particles and neutral beam ions, can resonate with shear Alfvén waves, causing the wave to grow via inverse Landau damping \citep{Rosenbluth_and_Rutherford_1975}. Additionally, Alfvén waves can grow via interactions with electron pressure gradients \citep{Heidbrink_2021}, ion temperature gradients \citep{Zonca_1999}, and ion temperature gradient-driven turbulence \citep{Riemann_2025}. In turn, the wave can deconfine the energetic particles \citep{Liu_Chen_1995}. The confinement of EPs is important for plasma heating and prevention of damage to reactor walls \citep{2008Heidbrink}. For improved EP confinement and Alfvénic stability, low or negative growth rates for modes at EP resonant frequencies are desirable.

The importance of Alfvénic stability has been demonstrated experimentally in stellarators. W7-AS showed increased MHD activity in the presence of energetic neutral beam ions, which likely caused the drive of the global Alfvén eigenmode (GAE) \citep{Weller_1994}. Likewise, TJ-II has shown increased MHD activity and Alfvénic chirping modes with the introduction of neutral beams \citep{TJII_Jimenez-Gomez_2011}. Today, proposed stellarators have much stronger energetic particle drive than existing experiments, which affects Alfvénic behavior. For example, Thea Energy plans to have two 10 MW neutral beam injectors in Eos \citep{Swanson_2025}. In addition, the proposed fusion pilot plant Infinity Two from Type One Energy is expected to have a volume-averaged $\beta_{\alpha}$ of $0.31\%$ \citep{type_one_infinity_two}.

To gain insight into Alfvénic stability, we turn to the shear Alfvén continuum.
Starting from reduced MHD \citep{Kruger_1998},
we can obtain an equation for the shear Alfvén continuum. First, we obtain a vorticity equation representing the wave
\begin{equation}
    B \nabla_{\parallel} \left ( \frac{\nabla_{\perp}^2(\nabla_{\parallel} \Phi)}{B} \right )
    + \frac{\omega^2}{v_A^2} \nabla_\perp^2 \Phi = 0,
    \label{eq:SAE}
\end{equation}
where $\omega$ is the frequency, $\Phi$ is the perturbed electrostatic potential, $v_A$ is the Alfvén velocity, $B$ is the magnetic field magnitude, and $\perp$ and $\parallel$ are with respect to the background magnetic field \citep{2021Sharapov}.
The Alfvén velocity is $v_A=B/\sqrt{\mu_0 \rho}$, where $\rho$ is the mass density and $\mu_0$ is the permeability of free space. This vorticity equation contains both continuous solutions that form shear Alfvén continua, such as the continua shown in Figures \ref{fig:continuum_and_gaps}a-e, and discrete gap solutions that fall within continuum gaps. The Alfvén velocity is $v_A=B/\sqrt{\mu_0 \rho}$, where $\rho$ is the mass density, $\mu_0$ is the permeability of free space, and $B$ is the magnetic field strength. As seen in the Alfvén velocity and \eqref{eq:SAE}, the frequencies vary radially with plasma density $\rho$ and magnetic field strength $B$.
Along the continuous structures, radial phase mixing gives rise to continuum damping.
The gaps, however, lack continuum damping and are therefore important for wave growth. 
When a radially global mode sits within a frequency gap and resonates with EPs, the wave amplitude can grow unimpeded by continuum damping. This makes the gap structure key for predicting Alfvénic stability.

We can take the continuum limit of \eqref{eq:SAE} to keep only the continuous solutions. In the continuum limit, we keep the terms with the highest order radial derivative, obtaining
\begin{equation}
    B \nabla_{\parallel} \left ( \frac{|\nabla \psi|^2}{B} \nabla_{\parallel} \Phi \right )
    + \frac{\omega^2|\nabla \psi|^2}{v_A^2} \Phi = 0,
    \label{eq:Continuum_eq}
\end{equation}
where $\psi$ is the flux label in magnetic flux coordinates \citep{2021Sharapov,salat_shear_2001}.
For an inhomogeneous plasma, the resulting frequencies form a radially continuous frequency spectrum known as the shear Alfvén continuum. 

The shear Alfvén continuum equation \eqref{eq:Continuum_eq} is an eigenvalue problem, where $\Phi$ is the eigenmode and $\omega^2$ is the eigenvalue. 
Different representations of the eigenmode $\Phi$ can be used. For example, \cite{Konies_and_Kleiber} employ B-splines, while Stellgap uses a Fourier representation in Boozer coordinates \citep{Stellgap}. An advantage of the Stellgap approach, which we use, is that it makes the parallel gradient easy to compute. However, a drawback of this approach is that the included Fourier modes must be prescribed by the user. Additionally, the mode selection suffers from a closure problem since it must be truncated. The choice of Fourier modes and their truncation is described in \secref{sec:mode_schemes}.
Using the Fourier basis in Boozer coordinates, the eigenmode $\Phi$ is discretized into a superposition of Fourier modes $m,n$, each with an amplitude $\Phi_{m,n}$:
\begin{equation}
    \Phi = \sum_{m,n} \Phi_{m,n} \cos(m \theta - n \zeta).
    \label{eq:fourier_decomp}
\end{equation}
The continuum is labeled according to its dominant Fourier mode, defined as the Fourier mode $m,n$ with the largest amplitude $\Phi_{m,n}$ in the corresponding eigenmode $\Phi$. 

When neglecting spatial coupling, the eigenmodes become Fourier modes. Two Fourier modes (e.g., $m$,$n$ and $m+\delta m$,$n+\delta n$) can have the same frequency and cross at some flux surface $s$, where $s$ is the normalized toroidal flux. With spatial coupling, a gap can form in a nonuniform magnetic field at this frequency. The formation of a gap at this frequency is analogous to a Bragg reflector \citep{2008Heidbrink}. In a Bragg reflector, there is a band of frequencies at which there is nearly total reflection and nearly no power transmission. The near zero power transmittance is caused by the destructive interference of the forward-propagating waves. For Alfvén waves, counter-propagating waves can destructively interfere, causing a gap in the shear Alfvén continuum. The gaps are identified by the difference in $m$ and $n$ of the crossing dominant Fourier modes (i.e., $\delta m,\delta n$). A closer look at the development of gaps in QS configurations using perturbation theory is presented in \citep{Paul_Hyder_Rodriguez_Jorge_Knyazev_2025}. 
The continuum gap structure is closely related to the flux-surface shaping \citep{Paul_Hyder_Rodriguez_Jorge_Knyazev_2025}, as seen by the effect of geometry on $|\nabla \psi|^2$ and $B$ in \eqref{eq:Continuum_eq}, meaning different equilibria exhibit different gap structures. In tokamaks, the continua are generally dominated by toroidal Alfvén eigenmode (TAE) and elliptical Alfvén eigenmode (EAE) gaps. In stellarators, we find that the continua are dominated by EAE, helical Alfvén eigenmode (HAE), and mirror Alfvén eigenmode (MAE) gaps with narrow TAE gaps. The width of EAE and HAE gaps is driven by rotating ellipticity, and the width of MAE gaps is driven by toroidal variation in the elongation \citep{Paul_Hyder_Rodriguez_Jorge_Knyazev_2025}. In \secref{sec:resonance_with_alphas}, we find that resonance with passing fusion-born alpha particles at the birth energy may be strong for Alfvén eigenmodes in the HAE(2,$N_{\text{fp}}$) gap of quasi-axisymmetric (QA) configurations, which is often the largest gap.

We build on \cite{Paul_Hyder_Rodriguez_Jorge_Knyazev_2025} by looking at the numerically produced continua under full equilibrium geometry. \cite{Paul_Hyder_Rodriguez_Jorge_Knyazev_2025} takes a near-axis approach to make analytical progress in improving our understanding of stellarator shear Alfvén continuum gap structure. We now look at the effects of the full geometry and shaping on the shear Alfvén continuum by studying several fusion pilot plant (FPP) configurations. Additionally, we go beyond QS configurations to include QI equilibria.

The flux-surface geometry can, in principle, be optimized to minimize continuum gaps and their widths, improving Alfvénic stability. A preliminary study using an analytical gap-width metric \citep{Paul_Hyder_Rodriguez_Jorge_Knyazev_2025} produced a Wistell-A-like \citep{Bader_Wistell-A} configuration with smaller high-frequency gap widths from weaker flux-surface shaping than Wistell-A. Future work could optimize directly against the shear Alfvén continuum gap structure.

Since the flux-surface shaping and continuum gap structure are closely related, a better understanding of the shear Alfvén continuum and resonant Alfvén eigenmodes across different classes of stellarators is required. An overarching comparison and study of shear Alfvén continua across modern stellarator equilibria has yet to be presented. 
Some comparisons are made between high- and low-aspect-ratio stellarators in the Stellgap paper \citep{Stellgap}, and Kolesnichenko \textit{et al.} compare the gap structure for different Helias configurations \citep{Kolesnichenko2001}.
However, much about the available equilibria today has changed since those two papers. For example, it is now possible to produce equilibria with much lower QS error, such as the Landreman-Paul precise QA and precise quasi-helically symmetric (QH) equilibria \citep{Landreman_paul_preciseQA}, and produce next-generation QI equilibria optimized for lower plasma turbulence, called SQuIDs \citep{Goodman_SQuID_2024}. Moreover, a large number of QS and QI equilibria are available through the QUASR \citep{Giuliani_Rodriguez_Spivak_2025} and ConStellaration databases \citep{cadena2025constellaration}, respectively.
In addition to the progress made in stellarator and fusion pilot plant (FPP) design and optimization, there is a need for increased Alfvénic stability analysis in new FPP designs. For example, Type One Energy spent considerable effort analyzing the Infinity Two stellarator for Alfvénic losses \citep{type_one_infinity_two}. Understanding the shear Alfvén continuum and the geometry that affects it is a strong first step in the Alfvénic stability analysis of new FPPs coming from the private sector. 

Due to the combination of large availability of well optimized QI and QS equilibria and the need from private-industry, it is now worth comparing the shear Alfvén continuum of different equilibria. 
Stellgap \citep{Stellgap} is used to calculate the shear Alfvén continuum across five stellarator configurations: Landreman $\beta=2.5\%$ QA \citep{2022Landremanb}, Landreman $\beta=2.5\%$ QH \citep{2022Landremanb}, ARIES-CS QA \citep{Ku2008-ARIES-CS}, Ku-Boozer QH \citep{Ku_Boozer_2011}, and a high-shear Stable Quasi-isodynamic Design (SQuID) configuration \citep{Goodman_SQuID_2024}. 
The geometry of the configurations and their flux shaping are discussed in \secref{sec:results}. 

The considerations, tools developed, and results of studying the shear Alfvén continuum in QA, QH, and QI equilibria are presented in this paper.
A discussion of frequency normalization and justification for the frequency ranges used in this paper, and methods for selecting the continuum Fourier modes and a comparison of the mode selection schemes are presented in \secref{sec:numerics}. It is found that carefully chosen selective mode schemes result in similar continua to using a full block of modes.
A novel algorithm for identifying and labeling continuum gaps, gap widths, and crossings that become gaps (i.e., avoided crossings) is presented in \secref{sec:gap_detection}. The method automatically labels gaps and computes the widths of both the gaps and the individual avoided crossings that make up them. 
Comparing the gap structure of equilibria requires both the gap type (i.e., HAE, MAE, etc.) and the gap widths associated with them. The gap widths serve as a proxy for Alfvénic stability.
Using one of the selective mode schemes and the gap detection algorithm, the continua for the five equilibria are compared in \secref{sec:results}. The gap widths are compared against an analytical metric in \secref{sec:grad_psi2_overview}.
The resonance of passing fusion-born alpha particles at gap frequencies is assessed in \secref{sec:resonance_with_alphas}. It is found that all configurations have a HAE(2,$N_{\text{fp}}$) as the largest gap, but this gap is often not resonant with passing fusion-born alphas for QH and QI equilibria. This may cause QH and QI equilibria to be easier to optimize for lower Alfvén-driven alpha losses.

\section{Fourier mode selection}
\label{sec:numerics}

There are several methods for selecting the Fourier modes \eqref{eq:fourier_decomp} that should be included in the continuum calculation. The method used in this paper will take advantage of the continuum frequency bounds relevant to fusion-born alphas in FPPs. The frequency bounds used in the mode schemes are normalized. Normalization is reviewed in \secref{sec:frequency_normalization}, and the frequency bounds are discussed in \secref{sec:frequency_range}. Mode families and mode schemes are discussed in \secref{sec:mode_families} and \secref{sec:mode_schemes}, respectively.

\subsection{Frequency normalization}
\label{sec:frequency_normalization}

To motivate the need for a normalized frequency, we turn to the shear Alfvén dispersion relation in terms of the poloidal and toroidal Fourier mode numbers, $m$ and $n$. The shear Alfvén wave dispersion relation in a slab is $\omega^2 = k_\parallel^2 v_A^2$. To obtain the dispersion relation in terms of $m$ and $n$, we begin by finding expressions for $k_\parallel$ and $v_A$. We start with the near-axis approximation in Boozer coordinates \citep{1991Garren} to simplify $k_\parallel$ and $v_A$.

Under the near-axis expansion, the distance from the magnetic axis is defined as 
\begin{equation}
    r=\sqrt{2 \psi / B_0},
    \label{eq:distance_to_magnetic_axis}
\end{equation}
with $\psi$ as the toroidal flux label and $B_0$ as the magnetic field strength on the magnetic axis \citep{1991Garren}.
In this expansion, the magnetic field strength $B$, rotational transform $\iota$, and the toroidal and poloidal covariant components of the magnetic field, $G$ and $I$, are given by
\begin{align}
    B(r) &= B_0(1+r \overline{\eta} \cos{(\theta - N \zeta)}) + \mathcal{O}(r^2) \\
    \iota(r) &= \iota_0 + \mathcal{O}(r^2) \\
    G(r) &= G_0 + \mathcal{O}(r^2) \\ 
    I(r) &= \mathcal{O}(r^2), 
    \label{eq:near-axis_assumptions}
\end{align} 
where $\overline{\eta}$ is a constant that defines the magnetic field \citep{1991Garren} and $N$ is the helicity. In the following discussion, we will retain only the on-axis terms, which give $B = B_0$, $\iota = \iota_0$, $G = G_0$, and $I = 0$. 

Assuming wave solutions $\Phi \sim e^{i(m \theta - n \zeta)}$, the parallel wavenumber is $k_{\parallel} \approx \mathbf{\hat{b}} \cdot \nabla \approx (\iota m - n) B / (G + \iota I)$, which under the on-axis assumption, becomes 
\begin{equation}
    k_{\parallel} \approx (\iota_0 m - n) \frac{B_0}{G_0}  = \frac{\iota_0 m - n}{R_0},
    \label{eq:k_parallel}
\end{equation}
where $R_0 = G_0 / B_0$ is the effective major radius of the magnetic axis. The Alfvén frequency is $\omega_A = v_A / R$. Continuing with the on-axis assumption, the Alfvén velocity $v_A^0$ is $B_0/\sqrt{\mu_0 \rho_0}$ and the Alfvén frequency is
\begin{equation}
    \omega_A^0 = \frac{v_A^0}{R_0} = \frac{B_0^2}{G_0 \sqrt{\mu_0 \rho_0}}.
    \label{eq:Alfven_frequency_on_axis}
\end{equation}
Using \eqref{eq:k_parallel} and \eqref{eq:Alfven_frequency_on_axis}, the dispersion relation in terms of $m$, $n$, and $\omega_A^0$ is
\begin{equation}
    \omega_0^2 = (\iota_0 m - n)^2 (\omega_A^0)^2.
    \label{eq:dispersion_relation_on_axis}
\end{equation}
Eq. \eqref{eq:dispersion_relation_on_axis} naturally lends itself to the definition of a normalized frequency
\begin{equation}
    \overline{\omega}_0 \equiv \omega / \omega_A^0 .
    \label{eq:near-axis_normalized_freq}
\end{equation}
Additionally, the use of the normalized frequency makes the frequency bounds dimensionless, which allows the Fourier mode bounds in \secref{sec:mode_schemes} to be expressed in terms of dimensionless parameters.

In addition to its role in the design of mode schemes, a normalized frequency is useful for the gap detection method, as it allows the use of filters that require low radial shear to function properly (see \secref{sec:gap_detection} \gapref{d_gap_labeling}).

\subsection{Physically-motivated frequency range}
\label{sec:frequency_range}

The continuum equation \eqref{eq:Continuum_eq} is no longer valid once finite-Larmor-radius (FLR) effects become important.
So, we estimate an upper bound for the continua frequency range using FLR effects associated with fusion-born alphas.

From \eqref{eq:k_parallel}, \eqref{eq:dispersion_relation_on_axis}, and \eqref{eq:near-axis_normalized_freq},
\begin{align}
\left \{
\begin{array}{l}
    k_\parallel \sim (\iota_0 m - n) / R_0 \sim \overline{\omega}_0 / R_0 \\
    k_\perp \sim m/r
\end{array}
    \right. ,
\label{eq:ks}
\end{align} 
where $r$ is given in \eqref{eq:distance_to_magnetic_axis}.
When $k_\perp \rho_{\text{ep}} \sim 1$, where $\rho_{\text{ep}}$ is the energetic particle Larmor radius, FLR effects become important \citep{Gorelenkov_1999}, so we choose to resolve $k_\perp \lesssim \rho_{\text{ep}}^{-1}$. From reduced MHD, $k_\parallel/k_\perp \ll 1$ \citep{Kruger_1998} and 
\begin{align}
    \frac{k_\parallel}{k_\perp} \sim \frac{\overline{\omega}_0}{R_0 k_\perp} \gtrsim \overline{\omega}_0 \frac{\rho_{\text{ep}}}{R_0}.
\label{eq:k_par_over_k_perp}
\end{align}
Therefore, we choose to resolve $\overline{\omega}_0 \rho_{\text{ep}} / R \ll 1$. For a fusion pilot plant, such as ARIES-CS, $R / \rho_{\text{ep}} \sim 100$ \citep{Ku2008-ARIES-CS}, so this limits us to low frequencies ($\overline{\omega}_0 \ll 100$) for modes of interest. The upper frequency bound, $\overline{\omega}_0 < 3$, is chosen to be consistent with the continuum frequency range considered in the Infinity Two alpha-particle study \citep{type_one_infinity_two}. 

\subsection{Mode families}
\label{sec:mode_families}

Toroidal mode numbers $n$ can be grouped into independent mode families.
It is sufficient to include only one mode family when computing the continuum in Stellgap. 
Each mode family can be uniquely identified by its lowest non-negative toroidal mode $n_{\text{start}}$. For our comparisons, we choose the $n_{\text{start}}=1$ mode family. Comparing different mode families, we find that the continua are qualitatively similar for all configurations.

Which toroidal modes constitute a mode family depends on the basis function being used. For a complex exponential basis, a single mode family is $\{n | n \in (n_{\text{start}} + cN_{\text{fp}}) \forall c \in \mathbb{Z} \}$, where $N_{\text{fp}}$ is the field period. For a cosine basis, as used in Stellgap, a single mode family must also include the $-n_{\text{start}}$ series. So, a mode family is the union of $\{n | n \in (n_{\text{start}} + cN_{\text{fp}}) \forall c \in \mathbb{Z} \}$ and $\{n | n \in (-n_{\text{start}} + cN_{\text{fp}}) \forall c \in \mathbb{Z} \}$. As an example, if $n_{\text{start}}=1$ and $N_{\text{fp}}=4$, then our mode family set is $\{...-7,-3,1,5,9...\}$ for an exponential basis. And, if we are using a cosine basis, then our mode family set is $\{...-7,-5,-3,-1,1,3,5,7,9...\}$. Additionally, for the cosine basis, we have $1+(N_{\text{fp}}/2)$ independent mode families \citep{Schwab_1993}, where $1+(N_{\text{fp}}/2)$ is rounded down to the nearest integer.

\subsection{Mode choice schemes}
\label{sec:mode_schemes}

When computing the shear Alfvén continuum, the toroidal $n$ and poloidal $m$ modes included in the calculation, collectively referred to as the mode resolution or mode choice, must be carefully selected. The goal is to include as many modes as possible whose normalized frequency is less than the frequency upper bound (see \secref{sec:frequency_range}) at some flux surface $s$. Including only the modes that produce some frequencies below the upper bound reduces computational complexity and lowers the condition number of the eigenvalue problem, which is the ratio of the largest to the smallest eigenvalues. 
Eigenmodes whose frequencies are always outside the frequency range of interest can increase the condition number and negatively affect the numerical stability of the eigenvalue problem.

We assume the cosine basis used by Stellgap.
With the cosine basis, we can choose $m$ to be strictly positive without loss of generality. 
Also, the $m=0,n=0$ mode should be excluded, since in \eqref{eq:fourier_decomp} the corresponding term of $\Phi$ is just the constant $\Phi_{0,0}$; the term's parallel gradient vanishes, which would leave the eigenvalue problem with degenerate solutions.
For $m=0$, only $n>0$ is included, since the basis is even.

Lastly, before we discuss the different mode choice schemes, it is important to mention a closure problem that impacts all of them. The Fourier modes in a given mode family are infinite, but some reasonable upper limit must be placed for numerical calculations.
As a result of the required mode truncation, there will almost always be eigenfrequencies that exist within Stellgap continuum gaps; this will be further discussed in \secref{sec:avoided_crossings}.

The simplest mode scheme is a block of modes. For the cosine basis, a block includes all Fourier modes in our mode family with $m$ bounded between 0 and some upper cutoff, and $n$ bounded between a lower and upper cutoff. 

\begin{figure}
    \centering 
    \begin{subfigure}{0.495\textwidth}
        \includegraphics[width=\textwidth]{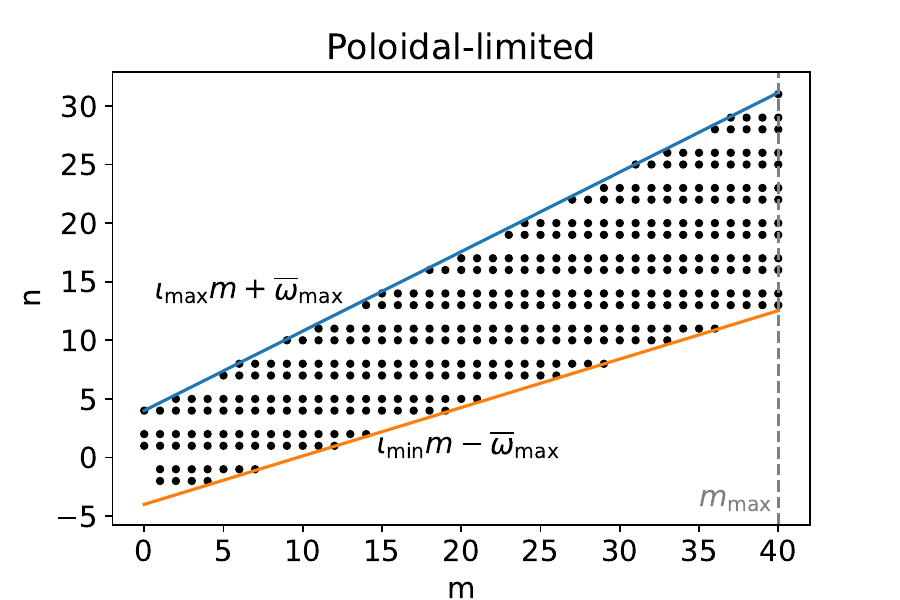}
        \caption{}
        \label{fig:poloidal-limited}
    \end{subfigure}
    \begin{subfigure}{0.495\textwidth}
        \includegraphics[width=\textwidth]{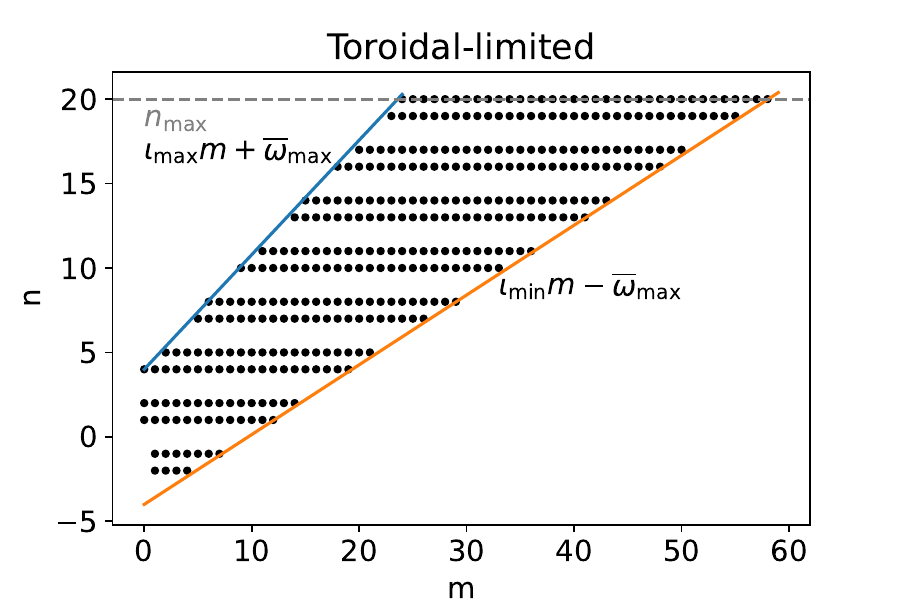}
        \caption{}
        \label{fig:toroidal-limited}
    \end{subfigure}
    \begin{subfigure}{0.495\textwidth}
        \includegraphics[width=\textwidth]{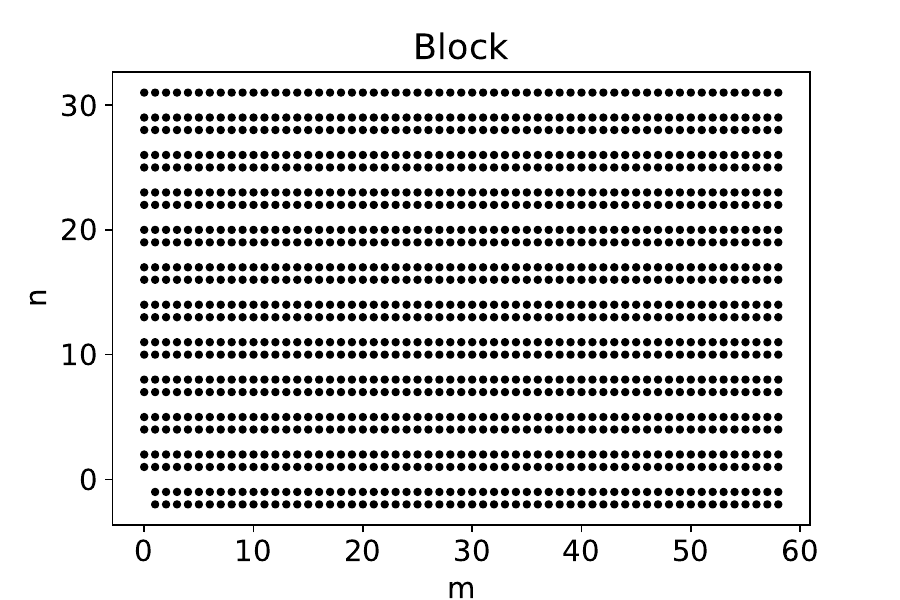}
        \caption{}
        \label{fig:block}
    \end{subfigure}
    
    \caption{Each figure contains dynamical modes selected for ARIES-CS with $N_{\text{fp}}=3$ \citep{Ku2008-ARIES-CS}. We use $\overline{\omega}_{\max} = 4$, and the chosen mode family is $n_{\text{start}} = 1$. The mode-limited schemes are figures (a) \textbf{poloidal-limited scheme:} limits the largest poloidal mode number allowed and (b) \textbf{toroidal-limited scheme:} limits the largest toroidal mode number allowed. Figure (c) is a \textbf{block} containing all the modes in figures (a) and (b). Figures (a) and (b) have a similar number of modes with 361 and 363, respectively. All the plots are the full mode family when using a cosine basis, like in Stellgap \citep{Stellgap}.}
    
    \label{fig:mode_choice_cone}
\end{figure}

Besides a block mode scheme, a selective mode scheme, such as Figures \ref{fig:mode_choice_cone}a,b, can be used. A selective mode choice scheme can be developed by taking advantage of the dispersion relation \eqref{eq:dispersion_relation_on_axis} and the frequency range of interest.
The frequency range for the QS and QI FPPs in our case is $\overline{\omega} \in [0,3]$ as mentioned in \secref{sec:frequency_range}. To ensure we include all the frequencies and Fourier modes that couple within [0,3], we will use a maximum normalized frequency, $\overline{\omega}_{\text{max}}$, of 4. From the normalized dispersion relation $\overline{\omega}_0^2 = (\iota_0 m - n)^2$, all Fourier modes that have frequencies below $\overline{\omega}_{\text{max}}$ satisfy $(\iota_0 m - n)^2 < \overline{\omega}_{\text{max}}^2$. When accounting for radially varying rotational transform $\iota(s)$, this condition becomes $(\iota(s) m - n)^2 < \overline{\omega}_{\text{max}}^2$. Without loss of generality, we can choose $\iota$ to be positive. Using the extrema of $\iota(s)$, $\iota_{\text{min}}$ and $\iota_{\text{max}}$, yields the following conditions on the toroidal mode $n$:
\begin{equation}
    \iota_{\min} m - \overline{\omega}_{\max} < n < \iota_{\min} m + \overline{\omega}_{\max}
    \label{eq:n_eq_from_iota_min}
\end{equation}
from $\iota_{\min}$ and
\begin{equation}
    \iota_{\max} m - \overline{\omega}_{\max} < n < \iota_{\max} m + \overline{\omega}_{\max}
    \label{eq:n_eq_from_iota_max}
\end{equation}
from $\iota_{\max}$.
Since $\iota(s)$ sits between $\iota_{\min}$ and $\iota_{\max}$, a mode will have a frequency in $[0,\overline{\omega}_{\max}]$ at some $s$ if it satisfies either condition, so we take the looser bounds for $n$:
\begin{equation}
    \iota_{\min} m - \overline{\omega}_{\max} < n < \iota_{\max} m + \overline{\omega}_{\max}.
    \label{eq:condition_on_n}
\end{equation}

There are two methods that can be used to place an upper limit on the selective mode choice. The first is to set a limit on the poloidal mode number (\figref{fig:mode_choice_cone}a), which will be subsequently called the poloidal-limited mode scheme, and the second is to include a limit on the toroidal mode number (\figref{fig:mode_choice_cone}b), which will be subsequently called the toroidal-limited mode scheme. The modes included for the two selective mode choice schemes are as follows
\begin{flalign}
\left \{
\begin{array}{l}
    \textbf{{Poloidal-limited scheme:}} \\
    m \in [0, m_{\text{max}}] \\ 
    n \in [\iota_{\text{min}}m - \overline{\omega}_{\max}, \iota_{\text{max}}m + \overline{\omega}_{\max}] \text{ for each } m\\
    \\
    \textbf{{Toroidal-limited scheme:}} \\
    m \in [\text{max}(0, (n - \overline{\omega}_{\max})/\iota_{\text{max}}),(n + \overline{\omega}_{\max})/\iota_{\text{min}}] \text{ for each } n\\ 
    n \in [- \overline{\omega}_{\max}, n_{\text{max}}]
\end{array}
    \right. ,
    \label{eq:limited-scheme-limits}
\end{flalign}
where $m_{\text{max}}$ and $n_{\text{max}}$ are determined \textit{ad hoc} by the user to adjust the mode resolution.

Within the frequency range of interest, the continuum for the two selective mode choice schemes should have a similar structure with the same overall gaps as the block mode choice. A difference in the gap structure between the mode schemes indicates that the chosen resolution is insufficient.

A block of modes can be chosen to include all the modes ($m$,$n$) in the selective mode schemes. Such a block serves as a good benchmark for the selective mode schemes. 
An example block of modes for the selective mode choices in \figref{fig:mode_choice_cone}a and \figref{fig:mode_choice_cone}b is illustrated in \figref{fig:mode_choice_cone}c. 

\begin{figure}
    \centering 
    \begin{subfigure}[b]{0.45\textwidth}
        \includegraphics[height=4.8cm]{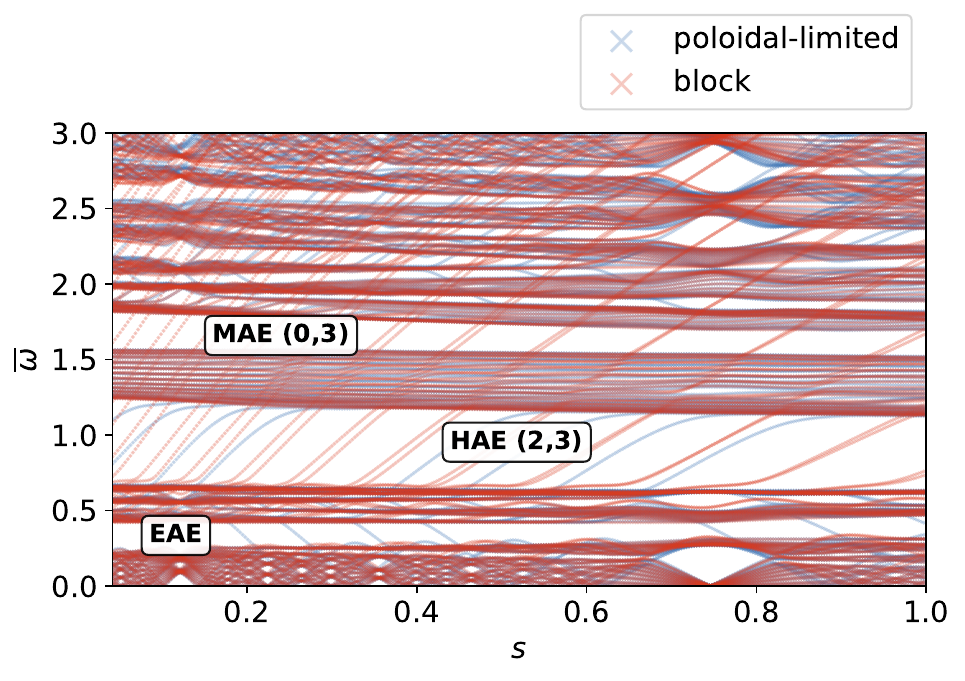}
        \caption{Block/poloidal-limited continua}
    \end{subfigure}
    \hfill
    \begin{subfigure}[b]{0.45\textwidth}
        \includegraphics[height=4.5cm]{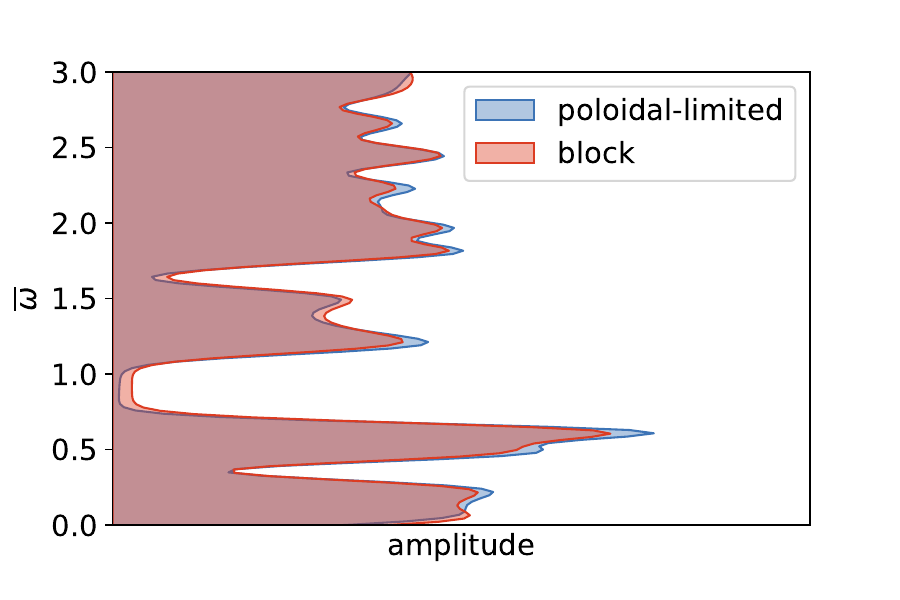}
        \caption{KDE of continuum points}
    \end{subfigure}
    \begin{subfigure}[b]{0.45\textwidth}
        \includegraphics[height=4.8cm]{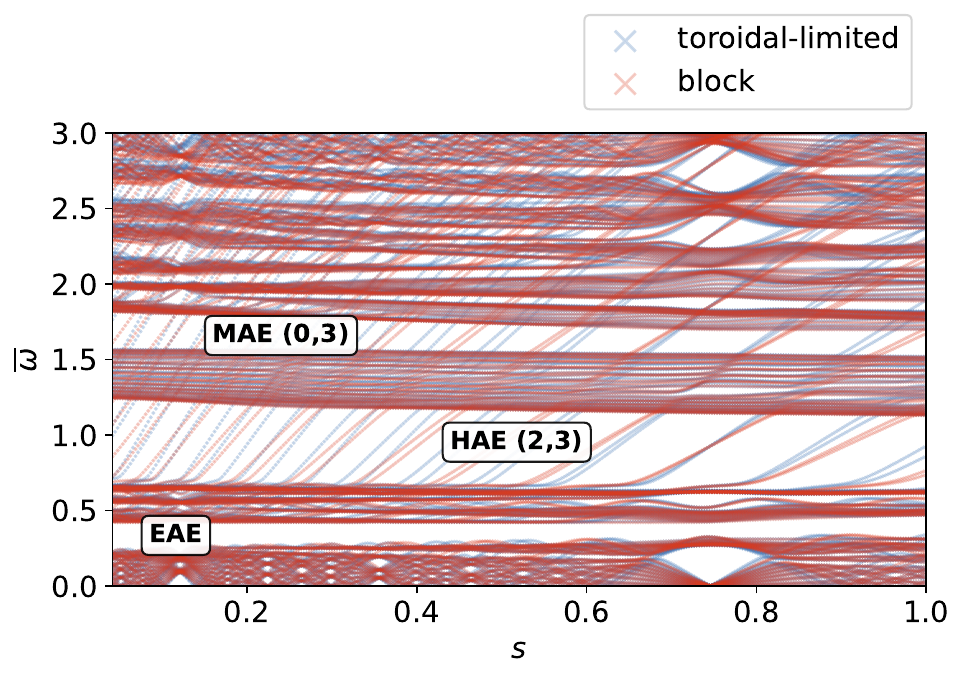}
        \caption{Block/toroidal-limited continua}
    \end{subfigure}
    \hfill
    \begin{subfigure}[b]{0.45\textwidth}
        \includegraphics[height=4.5cm]{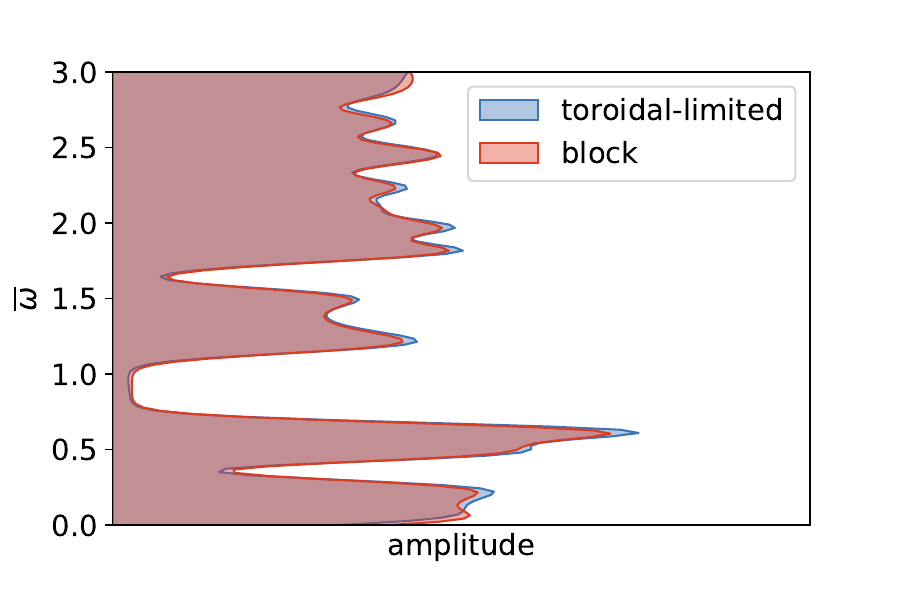}
        \caption{KDE of continuum points}
    \end{subfigure}
    
    \caption{The poloidal-limited and toroidal-limited mode schemes are benchmarked against the block scheme by overlaying their respective continua, and producing a kernel density estimation (KDE) of their continua. The KDE is that of the discrete continuum points ($s,\omega$) produced by Stellgap. The Fourier modes included are illustrated in \figref{fig:mode_choice_cone}.
    All continua are for the ARIES-CS \citep{Ku2008-ARIES-CS} configuration. The continuum gap structures show strong agreement across all three mode schemes.}
    \label{fig:gap_width_difference}
\end{figure}

To assess the different mode schemes, we overlap the continua of a poloidal-limited mode choice and a toroidal-limited mode choice, and compare their gap structure using a kernel density estimation (KDE) plot with an $\overline{\omega}$ bandwidth of 0.4. In \figref{fig:gap_width_difference}, the block scheme serves as the benchmark. It encompasses all the modes contained in the poloidal-limited and toroidal-limited mode schemes. The poloidal-limited and toroidal-limited schemes have roughly the same number of modes with 361 and 363 modes, respectively. As illustrated, the continua match very well, and the overall gap structure qualitatively agrees between all three mode schemes. The KDE plots include all frequencies, ignoring radial position. Low frequency density indicates a gap. The KDE plots match very well across all three mode schemes, indicating that the gap structure is in agreement. Since all three mode schemes benchmark well against each other, the poloidal-limited mode scheme or the toroidal-limited mode scheme should be chosen out of the three because they reduce computational load and avoid numerical stability problems associated with a large condition number. We choose the poloidal-limited scheme for this work.  

The poloidal-limited and toroidal-limited mode schemes can also be used beyond Stellgap and continuum calculations. The same selection of modes can be used in the eigenvalue solve of the shear Alfvén equation in AE3D and in FAR3D because it will include the modes that contribute frequencies to the relevant frequency range; see \citep{knyazev2026shearalfvenwaveinducedenergetic}.

Other mode schemes have been proposed in the literature. For example, Schwab employs a mode scheme that relies on the coupling between modes \citep{Schwab_1993}. As seen in \eqref{eq:Continuum_eq}, the coupling through the geometry for the continuum is given by $|\nabla \psi|^2$ and $B$. The Fourier decomposition of $|\nabla \psi|^2$ reveals which $\delta m, \delta n$ terms have the largest coefficients. 
Let us define a set $K$ containing the first few $\delta m, \delta n$ with the largest coefficients. To produce a collection of modes, one could choose a seed mode, an initial mode to start with. Then, generate all modes $m,n$ offset from the seed mode by all $\delta m, \delta n$ in $K$. The number of terms included in $K$ is an adjustable parameter, and it changes which gaps are targeted for inclusion in the continuum.
This process can then be repeated for each new mode until the resulting continuum is dense and has a sufficient number of $\omega = 0$ crossings. However, choosing the seed mode is non-trivial. Another caveat of this method is that it can include many modes outside of the frequency range relevant for the EPs of interest. So, we stick with the poloidal-limited schemes for this paper.

\section{Continuum gaps}
\label{sec:continuum_gaps}

Stellgap outputs all the eigenfrequencies at each flux surface by solving the eigenvalue problem in \eqref{eq:Continuum_eq}, given a radial resolution and a set of Fourier modes (see \eqref{eq:fourier_decomp} and \secref{sec:mode_schemes}). It also provides the dominant Fourier mode $m,n$ corresponding to each eigenfrequency, where the dominant mode has the largest amplitude $\Phi_{m,n}$ in the eigenmode $\Phi$ (see \eqref{eq:fourier_decomp}). We develop a method to automatically label continuum gaps and calculate their widths given frequency and mode data, such as that produced by Stellgap. Before discussing the algorithm, background on frequency crossings and the development of gaps are discussed in \secref{sec:avoided_crossings}. The algorithm is subsequently discussed in \secref{sec:gap_detection}.

\subsection{Crossings and avoided crossings}
\label{sec:avoided_crossings}

The normalized frequency and shear Alfvén dispersion relation are discussed in \secref{sec:frequency_normalization}. Both are used to develop insight into frequency crossings and gap formation. 

We first determine analytically the condition at which the frequencies of two dominant Fourier modes $m_i,n_i$ and $m_j,n_j$ (defined in \secref{sec:intro}) coincide. The normalized on-axis frequency for a particular Fourier mode is $\overline{\omega}_0^2 = (\iota_0 m - n)^2$ given by \eqref{eq:dispersion_relation_on_axis} and \eqref{eq:near-axis_normalized_freq}. 
With local $\iota (s)$ and $\rho (s)$, the normalized frequency is 
\begin{align}
    \overline{\omega} \equiv \omega / \omega_A(s),
    \label{eq:normalized_freq}
\end{align}
where $\omega_A = B_0^2 / (G_0 \sqrt{\mu_0 \rho (s)})$, and the local normalized frequency for a Fourier mode $m_i,n_i$ is $\overline{\omega}^2 = (\iota (s) m_i - n_i)^2$. The normalization removes the effects of mass density and rotational transform shear. The modes $m_i,n_i$ and $m_j,n_j$ then have the same frequency when
\begin{equation}
    \iota(s) m_i - n_i = \pm (\iota(s) m_j - n_j).
    \label{eq:eq_freqs}
\end{equation}
The $\pm$ in \eqref{eq:eq_freqs} indicates the possibility of co-propagating (positive sign) and counter-propagating (negative sign) waves, where only counter-propagating waves result in a gap. 
Due to the existence of gaps, we focus on the counter-propagating case in this paper.
A further discussion on the introduction of a gap for counter-propagating waves using perturbation theory is given in Section 4.1 of \citep{Paul_Hyder_Rodriguez_Jorge_Knyazev_2025}. 

\begin{figure}
    \centering 
    \includegraphics[width=0.8\textwidth]{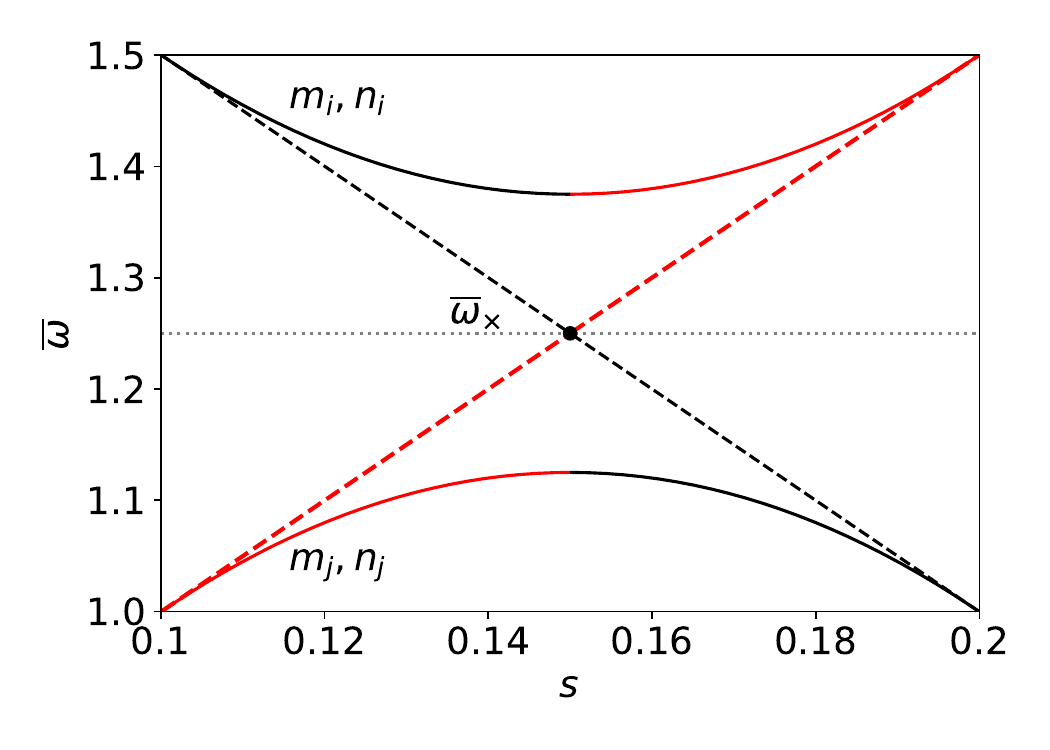}
    
    \caption{Depicted is an example of an avoided crossing between two modes $m_i,n_i$ and $m_j,n_j$ shown in black and red, respectively. The near-axis predicted frequency of the crossing is $\overline{\omega}_\times$. 
    The dashed lines show the scenario without spatial coupling, while the solid lines include it. The solid lines exhibit an avoided crossing.}
    
    \label{fig:crossing_figure}
\end{figure}

If we neglect spatial coupling in $|\nabla \psi|^2$ and $B$, no gaps form in the shear Alfvén continuum. This is illustrated by the crossing of the dashed lines in \figref{fig:crossing_figure}. The frequency at which the dominant modes match is
\begin{align}
    \overline{\omega}^2_\times = \left ( \frac{\iota(s) \Delta m - \Delta n }{2} \right )^2,
    \label{eq:freq_crosssing}
\end{align}
where $\Delta m = m_i - m_j$ and $\Delta n = n_i - n_j$.
This frequency will be used as a source of comparison in \secref{sec:results} for the Stellgap numerical continuum results against analytical expectations. With spatial coupling in $|\nabla \psi|^2$ and $B$, the crossing of counter-propagating waves, such as in \figref{fig:crossing_figure}, opens up to become an avoided crossing (solid lines) formed by two Fourier modes: $m_i,n_i$ (black) and $m_j,n_j$ (blue). Avoided crossings are labeled by $\Delta m, \Delta n$.

A collection of avoided crossings with the same $\Delta m, \Delta n$ forms a gap identified as the $\delta m,\delta n$ gap, where $\delta m=\Delta m, \delta n=\Delta n$. For example, all the $(\Delta m = 2, \Delta n = N_{\text{fp}})$ avoided crossings together form the helical Alfvén eigenmode (HAE) ($\delta m = 2, \delta n = N_{\text{fp}}$) gap. From now on, capital $\Delta$ is used when referring to avoided crossings and $\delta$ is used when referring to gaps. The nomenclature for the gaps is specified in \tabref{tab:AE_types}, and the gap names refer to the global eigenmodes that exist in continuum gaps.
The avoided crossings for a single $\delta m, \delta n$ gap should appear at various flux surfaces $s$ that span some significant radial region.

As discussed in \secref{sec:mode_schemes}, eigenfrequencies may exist within continuum gaps because of a numerical closure problem with the selected Fourier modes. This happens when a Fourier mode $m + \Delta m$, $n + \Delta n$ above the imposed poloidal or toroidal mode limit couples with a mode $m$, $n$ within the limit to form a $\Delta m$, $\Delta n$ avoided crossing. The removal of the $m + \Delta m$, $n + \Delta n$ mode can cause the $m$, $n$ mode to have frequencies within a gap. These spurious crossings through gaps are artifacts of solving the numerical problem rather than being of physical significance.

\begin{table}
    \centering
    \begin{tabular}{lcc}
        \textbf{Name} & \textbf{$|\delta m|$} & \textbf{$|\delta n|$} \\
        Global Alfvén Eigenmode (GAE) & 0 & 0 \\
        Toroidal Alfvén Eigenmode (TAE) & 1 & 0 \\
        Elliptical Alfvén Eigenmode (EAE) & 2 & 0 \\
        Non-circular Alfvén Eigenmode (NAE) & $\ge3$ & 0 \\
        Mirror Alfvén Eigenmode (MAE) & 0 & $\ge1$ \\
        Helical Alfvén Eigenmode (HAE) & $\ge1$ & $\ge1$
    \end{tabular}
    \caption{Nomenclature for Alfvén eigenmodes \citep{Stellgap}.}
    \label{tab:AE_types}
\end{table}

\subsection{Gap detection}
\label{sec:gap_detection}

To aid in the analysis of continua, we would like to automatically identify and label continuum gaps and calculate their average gap widths. Both needs are fulfilled in the following method.
To accomplish this, we need to parse the Stellgap output. Stellgap solves the shear Alfvén continuum eigenvalue problem of \eqref{eq:Continuum_eq} on a uniform grid in the normalized toroidal flux $s$, outputting the eigenfrequencies and their corresponding continuum eigenmodes. We also modify Stellgap to output the eigenvector for each eigenfrequency.

We can make some assumptions about the frequencies corresponding to a single dominant Fourier mode based on Sections \ref{sec:frequency_normalization} and \ref{sec:avoided_crossings}. From $\overline{\omega}^2 = (\iota (s) m_i - n_i)^2$ (see \eqref{eq:dispersion_relation_on_axis} and \eqref{eq:normalized_freq}), we expect analytically that the frequencies for a given dominant mode $m,n$ will follow continuous curves if $\iota (s)$ is continuous and we ignore spatial coupling in $|\nabla \psi|^2$ and $B$. With spatial coupling, we expect there to be, at avoided crossings, a switch in the mode that belongs to a continuous curve, as illustrated in \figref{fig:crossing_figure}, where the top solid curve switches from $m_i,n_i$ to $m_j,n_j$ and the bottom solid curve does the opposite. We also expect that for every switch from $m_i,n_i$ to $m_j,n_j$ there is an equivalent switch from $m_j,n_j$ to $m_i,n_i$ and the two together form an avoided crossing. 

For the gap detection method, we take advantage of the property that the eigenfrequencies and eigenvectors of the eigenvalue problem \eqref{eq:Continuum_eq} vary smoothly across flux surfaces. In \citep{Greenbaum_2020}, it is shown that the eigenvalues and eigenvectors of a square matrix with smoothly varying inputs result in smoothly varying eigenvalues and eigenvectors. The Stellgap matrices are always square and, when sound coupling is not included, positive definite and real. With regard to inputs, both $B$ and $|\nabla \psi|^2$ vary smoothly across flux surfaces. In the case of sound coupling, it is expected that the temperature and poloidal and toroidal currents also vary smoothly across flux surfaces. So, we can assume the eigenvalues and eigenvectors for the shear Alfvén continuum eigenvalue problem to vary smoothly across flux surfaces. Additionally, the eigenvalues and eigenvectors are differentiable \citep{Murthy1988,Giles2008}.

\begin{figure}
    \centering 
    \begin{subfigure}{0.49\textwidth}
        \includegraphics[width=\textwidth]{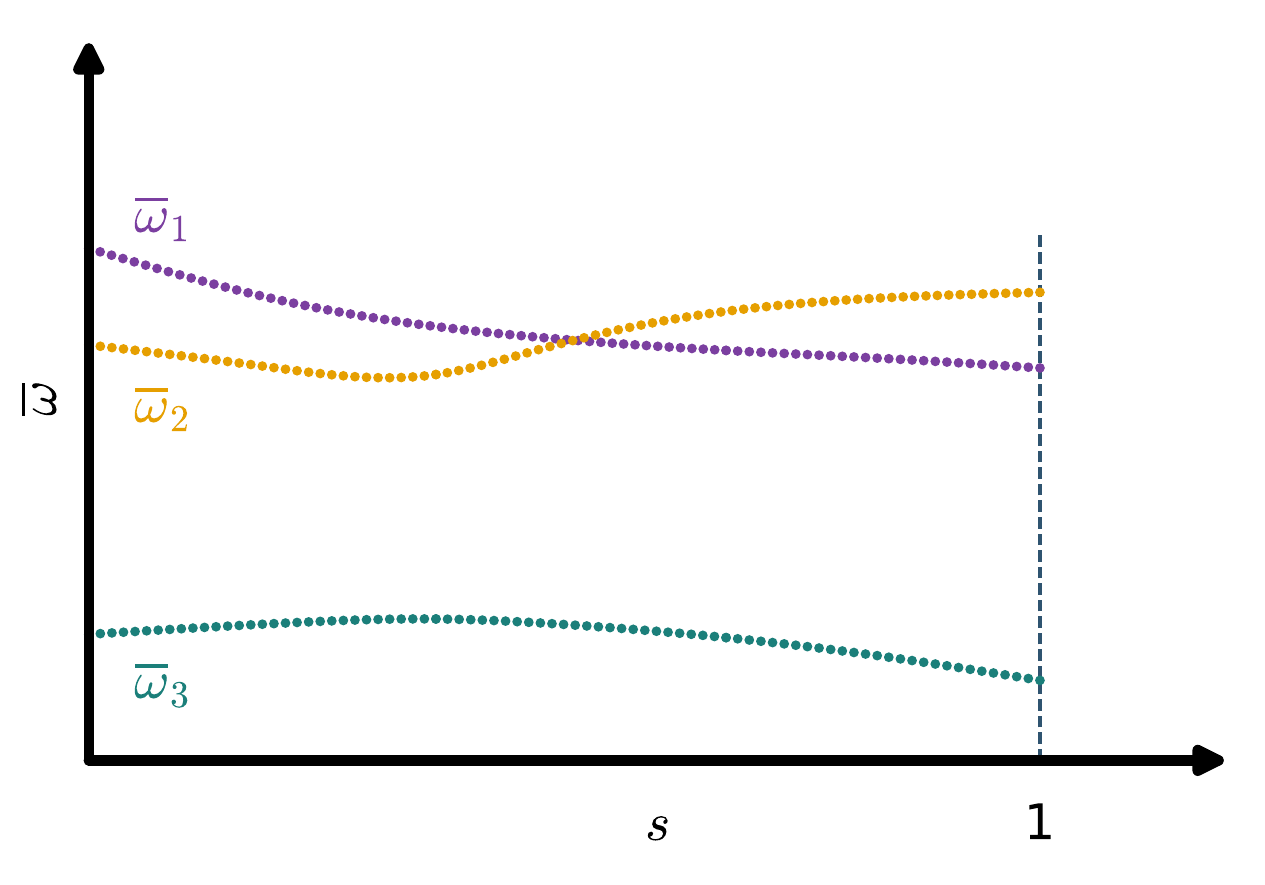}
        \caption{Track eigenvalues}
        \label{fig:step_a}
    \end{subfigure}
    \centering 
    \begin{subfigure}{0.49\textwidth}
        \includegraphics[width=\textwidth]{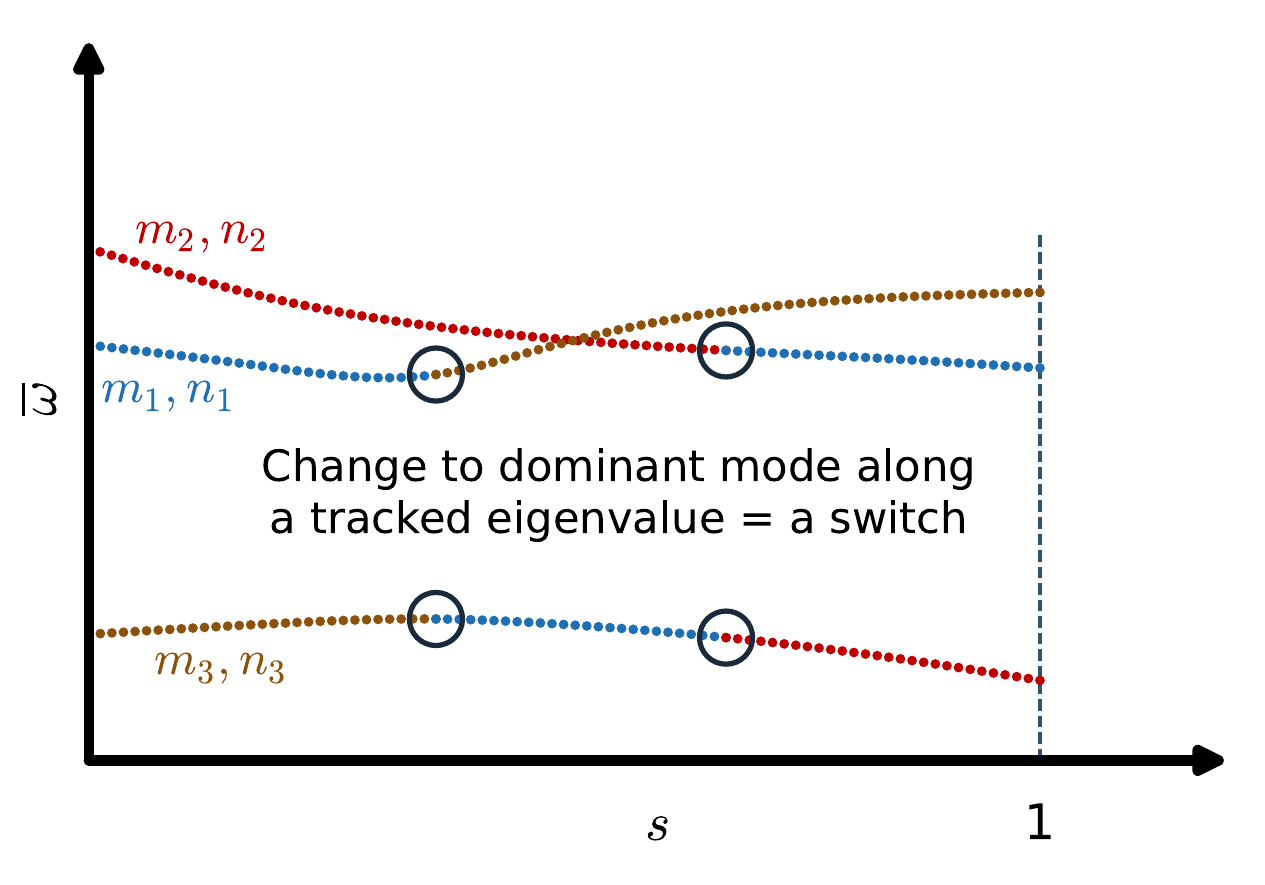}
        \caption{Find switches}
        \label{fig:step_b}
    \end{subfigure}
    \centering 
    \begin{subfigure}{0.49\textwidth}
        \includegraphics[width=\textwidth]{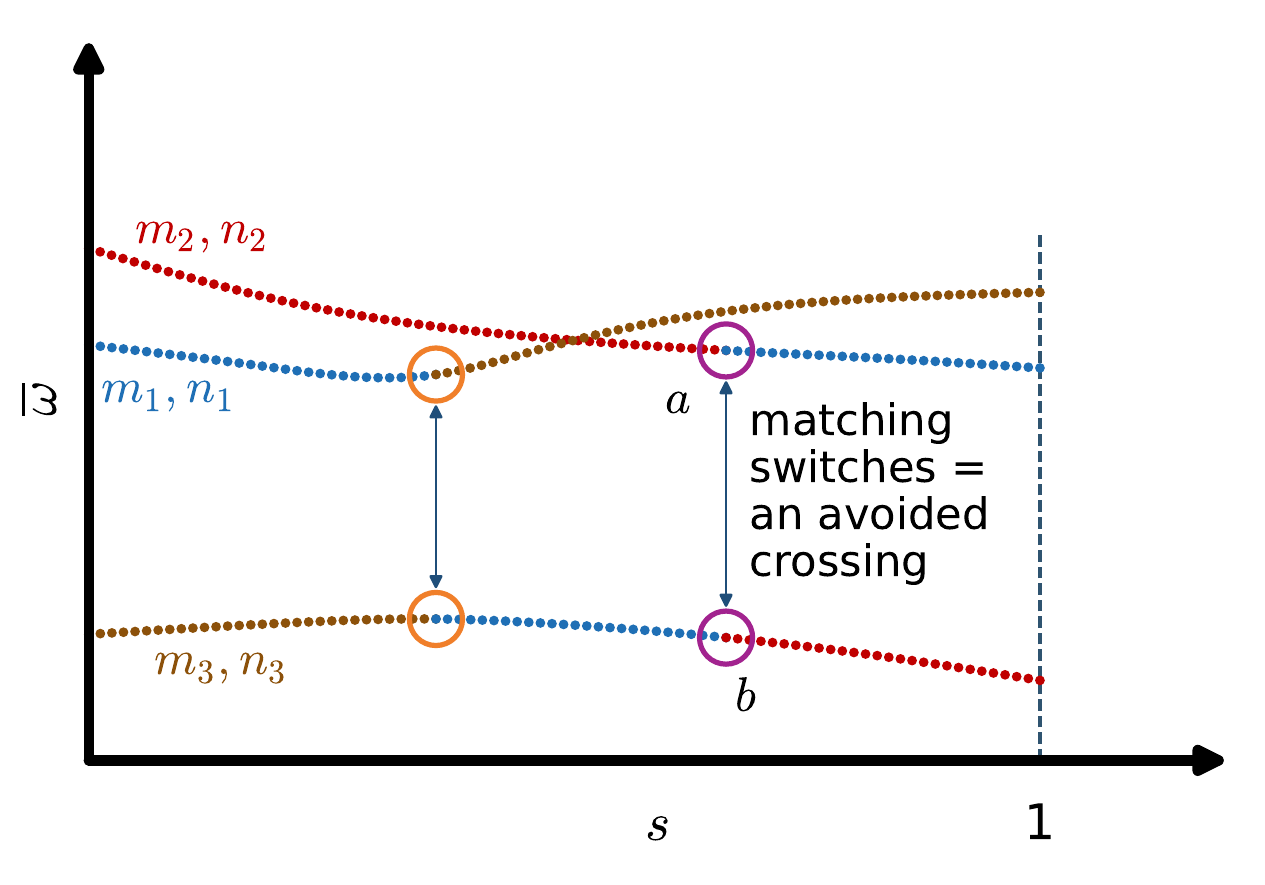}
        \caption{Find avoided crossings}
        \label{fig:step_c}
    \end{subfigure}

    \caption{The first three steps of the gap detection method are illustrated. (a) The first step is to separate all the eigenfrequencies into tracked eigenvalues based on the eigenvectors. (b) The second step is to find switches in dominant mode number along each tracked eigenvalue. (c) The third step is to pair these switches together in order to identify an avoided crossing. The switches are paired based on having the same modes $m_1,m_2,n_1,n_2$, but in opposite order.}
    \label{fig:gap_detection}
\end{figure}

We refer to a set of eigenfrequencies whose eigenvectors vary smoothly from one flux surface to the next as a tracked eigenvalue; each tracked eigenvalue has exactly one eigenfrequency on any given flux surface. We exploit this smoothness to develop a gap detection method. First, we sort the eigenfrequencies into curves, one for each tracked eigenvalue, which we call continuum curves. Next, we look for changes in the dominant mode number along each continuum curve. Each element of an eigenvector corresponds to the amplitude of a Fourier mode $m,n$, and the element with the largest magnitude identifies the dominant mode. When the dominant mode of a tracked eigenvalue changes between two adjacent flux surfaces, we call this a switch. For example, the upper continuum curve of \figref{fig:gap_detection}b switches from $m_2,n_2$ to $m_1,n_1$, and its matching switch is on the lower curve. A pair of matching switches identifies an avoided crossing.

Ideally, each dominant Fourier mode exists at no more than one frequency on each flux surface, forming continuous curves everywhere except at avoided crossings. Numerically, it is possible for several terms in the eigenmode Fourier decomposition to have similar magnitudes, which makes it difficult to find a clear dominant mode. The result is that multiple frequencies at a single flux surface can have the same dominant Fourier mode, 
and there is no guarantee that the eigenfrequencies of the same dominant Fourier mode will be part of the same continuum curve in the absence of an avoided crossing.

In order to systematically identify the gaps, avoided crossings are grouped by the same $\Delta m, \Delta n$. If at least $d$ avoided crossings of the same $\Delta m, \Delta n$ exist, the gap is labeled. There are some exceptions, which are in the discussion of filters below. For the mode resolution used in \secref{sec:results}, $d$ is chosen to be 3.
The number of avoided crossings that constitute a gap $d$ is a function of the mode resolution. If the mode resolution is higher, then $d$ can be higher. This method also works for gaps that do not radially persist through all flux surfaces as long as the mode resolution is high enough.

In short, the identification and labeling of Stellgap continuum gaps follows a five-step algorithm. Some steps are discussed in more detail in \appref{app:appendix_a}.

\begin{gap_detect}
\gapitem{\textbf{a)} }{a_normalization}{\textbf{Normalize the continuum}}\textbf{:} Normalizing the continuum removes large slopes caused by mass density $\rho(s)$ shear, which can otherwise produce significant radial variation in frequency. If trying to only identify avoided crossings, normalization is not necessary because steps (b)-(d) are not affected by strong slopes in the continuum. However, the first filter of step (e) assumes a normalized frequency. As seen in \eqref{eq:normalized_freq}, the normalized frequency used throughout this algorithm, $\overline{\omega}$, is $\omega / \omega_A(s)$, where $\omega_A(s)$ is the local Alfvén frequency. By normalizing in this way, configurations with large mass density shear can still use the first gap filter. If an unnormalized continuum is desired, the filter can still be utilized by normalizing the continuum, applying the filter, then unnormalizing the continuum.

\gapitem{\textbf{b)} }{a_forming_segments}{\textbf{Identifying eigenvalues by their eigenvectors}}\textbf{ (\figref{fig:gap_detection}a):} We track each eigenvalue across flux surfaces using its eigenvector, which varies smoothly from one surface to the next. We start with the number of tracked eigenvalues equaling the number of Fourier modes on our first flux surface. 
The eigenvectors for all the eigenfrequencies are saved and normalized into unit vectors. Moving from one surface to the next, we compute the inner product of each eigenvector on the current surface with every eigenvector on the previous surface, and pair those with the largest inner product as the same tracked eigenvalue. The pairing is one-to-one, so each tracked eigenvalue appears once on every flux surface. 
Each tracked eigenvalue forms a continuum curve.

\gapitem{\textbf{c)} }{b_finding_switches}{\textbf{Find switches}}\textbf{ (\figref{fig:gap_detection}b):} In the previous step, eigenvalue and eigenvector information at each flux surface is used to break the eigenfrequencies into continuum curves. Along each continuum curve, the dominant mode can change. In this step, we traverse radially through the frequencies of each tracked eigenvalue and save the locations where the dominant mode switches. The frequency and radial locations of where the dominant mode switches are saved. This switch is assumed to be either the top or bottom of an avoided crossing. The change in modes from $m_1,n_1$ to $m_2,n_2$ is also saved for each switch. At the end of this step, the frequency, flux surface, starting mode, and ending mode are known for all the switches.

\gapitem{\textbf{d)} }{c_find_avoided_crossings}{\textbf{Find avoided crossings}}\textbf{ (\figref{fig:gap_detection}c):} The third step is to identify individual avoided crossings. 
The third step is done by finding, for each switch identified in step \gapref{b_finding_switches}, a match with the same Fourier mode numbers and the same $\Delta m$, $\Delta n$, but with opposite sign. The opposite sign for $\Delta m$, $\Delta n$ indicates that, moving radially outward, the segment switches from $m_1, n_1$ to $m_2, n_2$ at one switch in the matching pair, and from $m_2, n_2$ to $m_1, n_1$ at the other. For example, if the first switch is labeled $a$ and the second switch is labeled $b$, as illustrated in \figref{fig:gap_detection}c, and the Fourier mode to the left of the switch is labeled $1$ and the mode to the right of the switch is labeled $2$, then $m_{1a}=m_{2b}$,$n_{1a}=n_{2b}$,$m_{2a}=m_{1b}$, and $n_{2a}=n_{1b}$ would be a match. If there are multiple switches that satisfy this matching criterion for $a$, the switch closest in $s$ is matched to $a$. If multiple switches are equally close in $s$ to $a$, the one closer in frequency is matched. Each switch can only have one match. A maximum threshold on the radial distance between two switches is also placed as $s_{\mathrm{threshold}}$, which we set to 0.1 by default. To avoid matching switches along the same continuum curve, we require that matching switches come from different tracked eigenvalues.

Not every switch has a matching switch whose modes change in the opposite direction, so some switches are left unpaired. Sometimes this is caused by multiple elements in a eigenvector having similar magnitudes, where the dominant mode is difficult to distinguish.

\gapitem{\textbf{e)} }{d_gap_labeling}{\textbf{Identify and label gaps}}\textbf{:} The fourth step is to identify and label gaps. First, we apply a filter that checks for and removes certain irregular avoided crossings. We check that each $\Delta m$, $\Delta n$ avoided crossing is near the center of the $\delta m$, $\delta n$ gap by verifying that its center lies within a gap width $\Delta \overline{\omega}_{\delta m, \delta n}$ of any other avoided crossing sharing the same $\Delta m$, $\Delta n$.
If that condition is not satisfied, the avoided crossing is considered irregular. This first filter should not be used when sound coupling is included, because the frequency shift of the beta-induced Alfvén eigenmode (BAE) may introduce a shear to continuum gaps that depends on the temperature profile. After that filter, we check if there is a sufficient number of avoided crossings. The minimum number of avoided crossings to be identified as a gap $d$ is based on the mode resolution used; here, it is chosen to be 3.
If there is a sufficient number of avoided crossings, a label is added to the gap using $\delta m,\delta n$ and \tabref{tab:AE_types}.

\end{gap_detect}

The five steps outlined above involve some additional details, which are elaborated on in \appref{app:appendix_a} starting with \gapref{a_forming_segments}. As part of the method, the width for each avoided crossing is calculated, and the average of them, excluding irregular avoided crossings, is the gap width.

With this gap-identification implementation, nearly all avoided crossings can be identified and labeled if desired. A more common use is to identify and label all gaps with a significant gap width. 

\section{Results and Discussion}
\label{sec:results}

\begin{figure}
    \centering 
    \begin{subfigure}[t]{0.48\textwidth}
        \includegraphics[width=\textwidth]{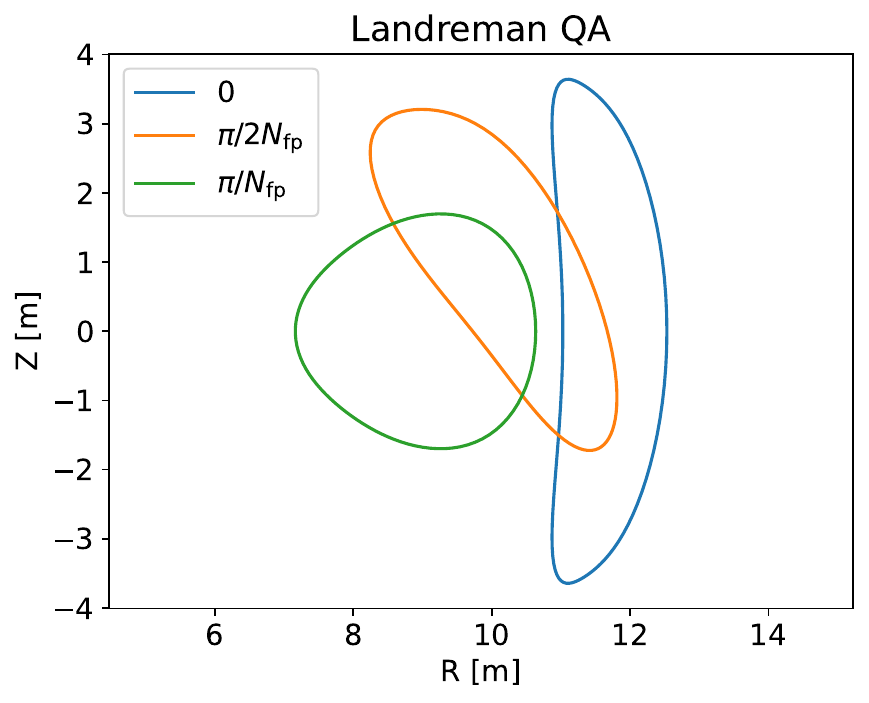}
        \caption{}
    \end{subfigure}
    \begin{subfigure}[t]{0.48\textwidth}
        \includegraphics[width=\textwidth]{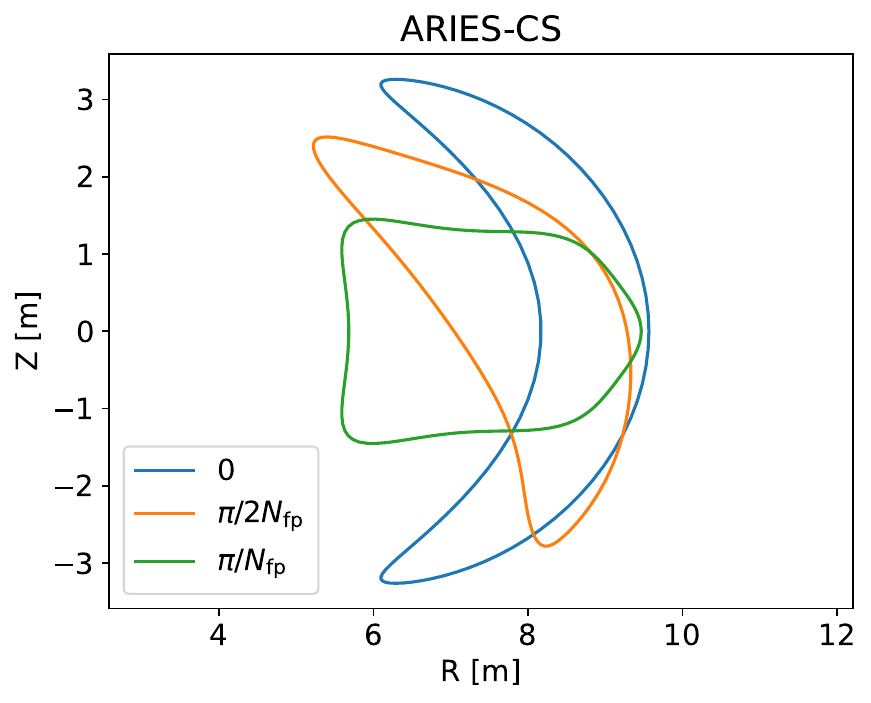}
        \caption{}
    \end{subfigure}
    \begin{subfigure}[t]{0.48\textwidth}
        \includegraphics[width=\textwidth]{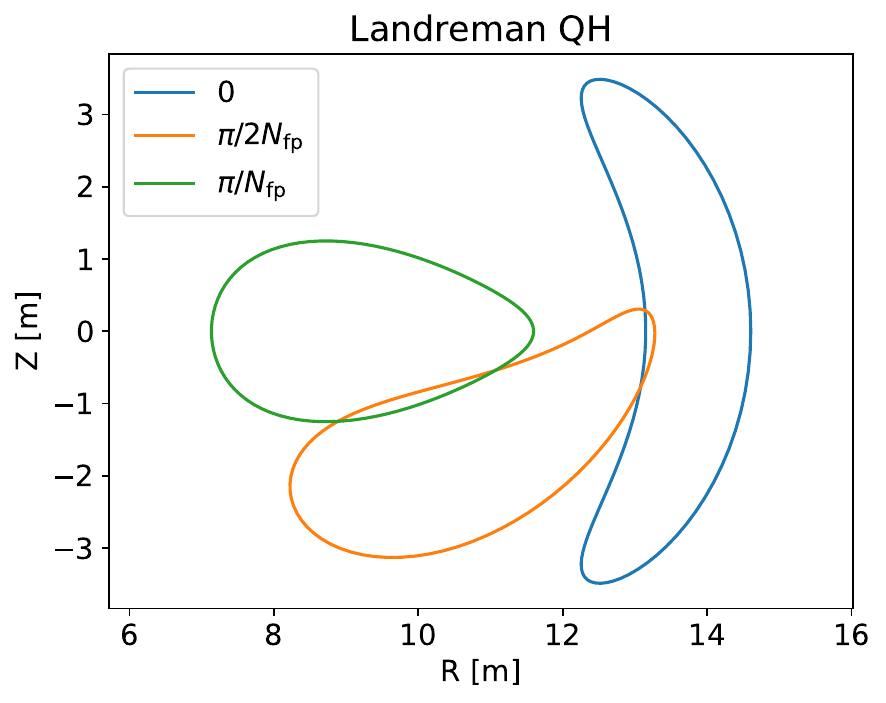}
        \caption{}
    \end{subfigure}
    \begin{subfigure}[t]{0.48\textwidth}
        \includegraphics[width=\textwidth]{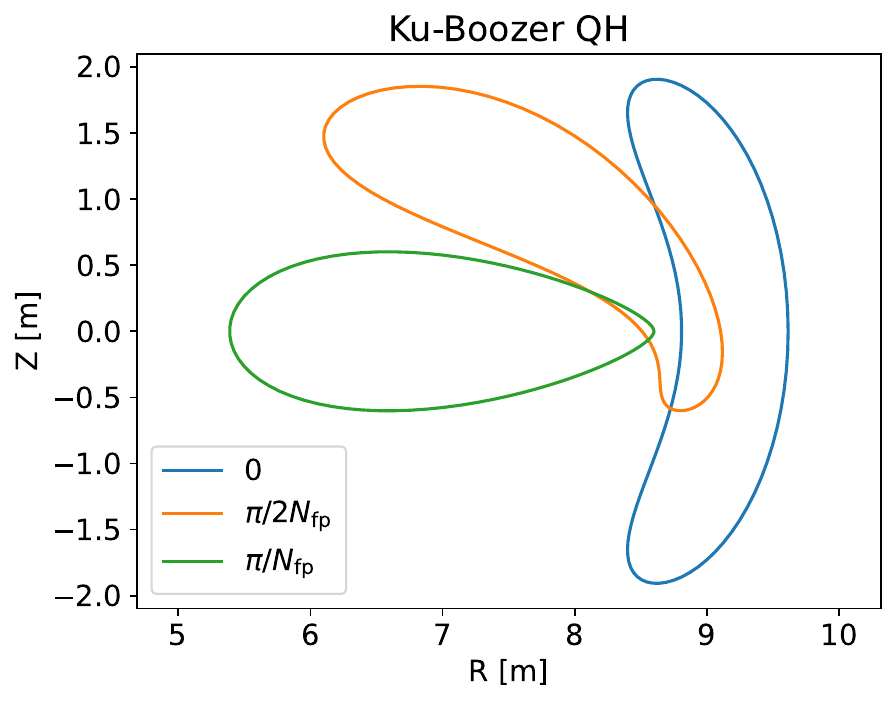}
        \caption{}
    \end{subfigure}
    \begin{subfigure}[t]{0.48\textwidth}
        \includegraphics[width=\textwidth]{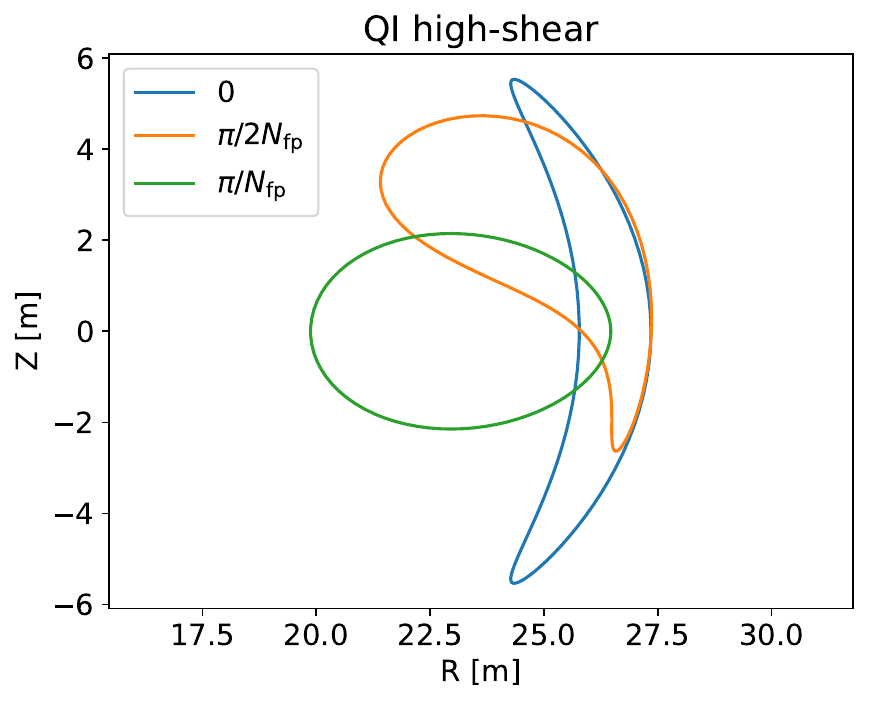}
        \caption{}
    \end{subfigure}
    \begin{subfigure}[t]{0.49\textwidth}
        \includegraphics[width=\textwidth]{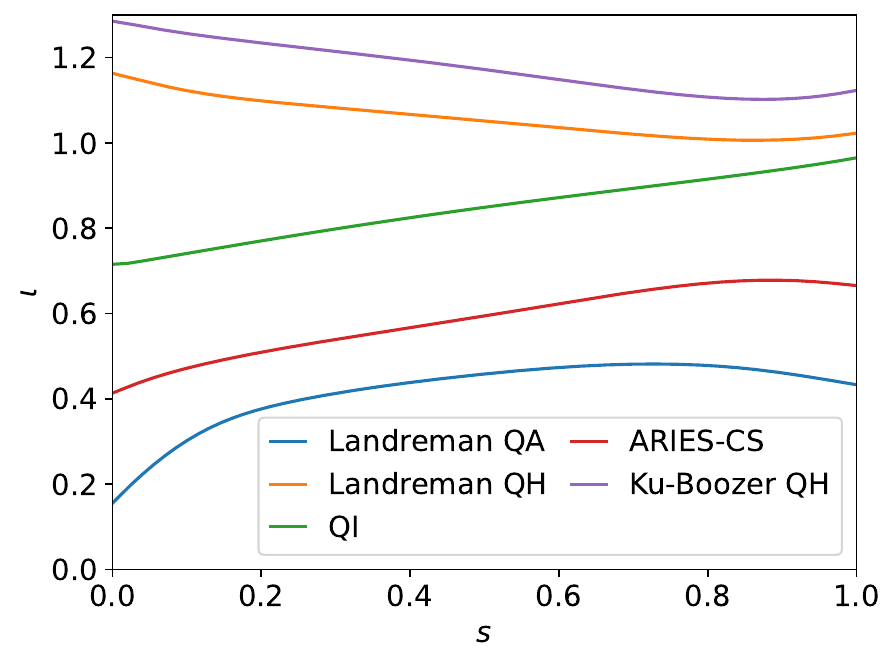}
        \caption{}
    \end{subfigure}
    \caption{Illustrated are the poloidal cross-sections at the last closed flux surface for toroidal angles $0$, $\pi/2$, and $\pi$ within a field period. The cross-section is taken normal to the magnetic axis. The configurations are as follows: (a) Landreman $\beta=2.5\%$ QA \citep{2022Landremanb}, (b) ARIES-CS QA \citep{Ku2008-ARIES-CS}, (c) $\beta=2.5\%$ Landreman QH \citep{2022Landremanb}, (d) Ku-Boozer QH \citep{Ku_Boozer_2011}.}
    \label{fig:cross-sections}
\end{figure}

To examine how geometry affects the shear Alfvén continuum structure and gap widths, we compute the continuum for several equilibria using Stellgap \citep{Stellgap}. We compare five reactor-scale configurations: two QA, two QH, and one QI 
equilibrium (see \secref{sec:intro}). The cross-section and $\iota$ profile for these five configurations are given in \figref{fig:cross-sections}.

\begin{figure}
    \centering 
    \begin{subfigure}[t]{0.49\textwidth}
        \includegraphics[width=\textwidth]{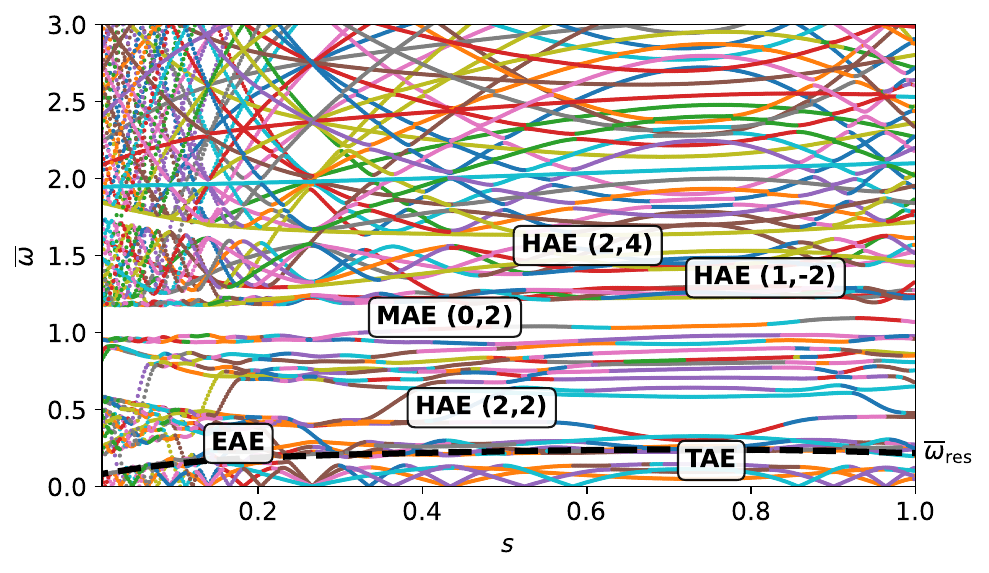}
        \caption{Landreman $\beta=2.5\%$ QA \citep{2022Landremanb}}
        \label{fig:QA_continuum}
    \end{subfigure}
    \hfill
    \begin{subfigure}[t]{0.49\textwidth}
        \includegraphics[width=\textwidth]{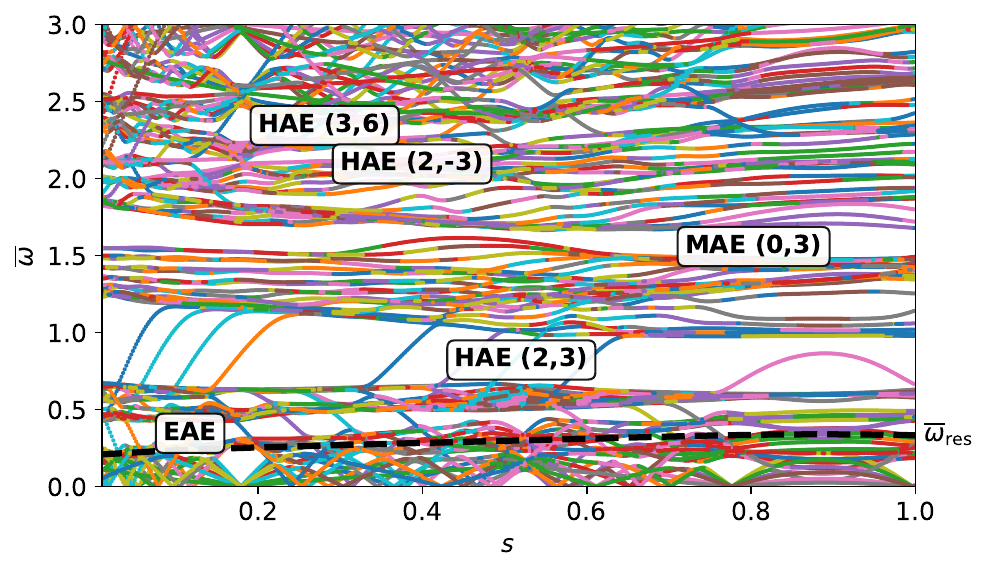}
        \caption{ARIES-CS QA \citep{Ku2008-ARIES-CS}}
        \label{fig:ARIES_continuum}
    \end{subfigure}

    \begin{subfigure}[t]{0.857\textwidth}
        \includegraphics[width=\textwidth]{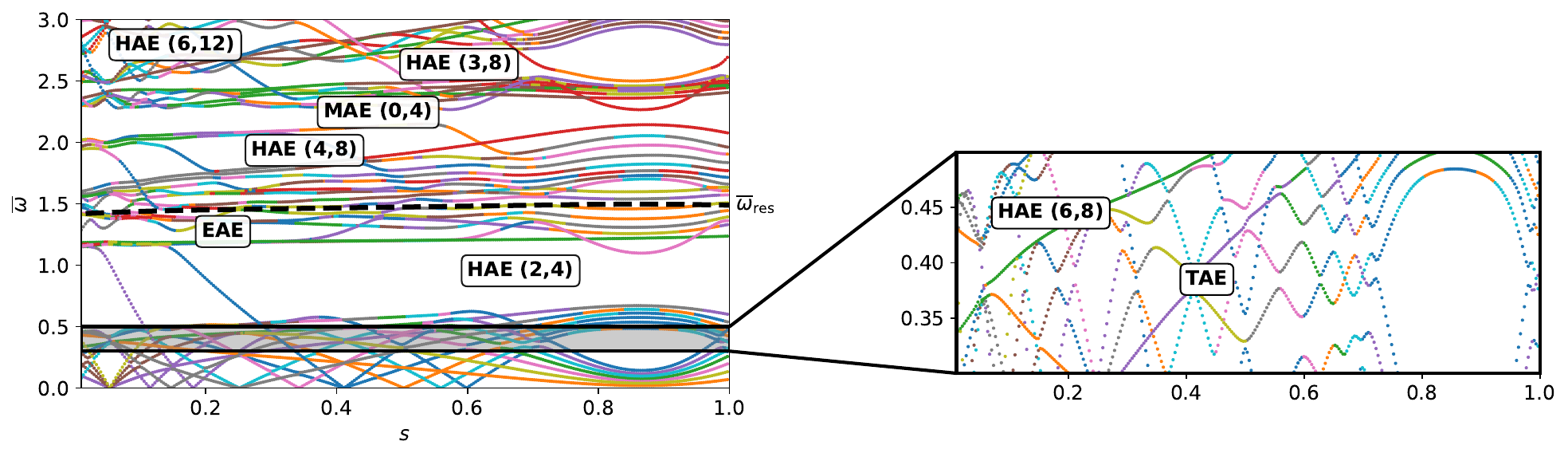}
        \caption{Landreman $\beta=2.5\%$ QH \citep{2022Landremanb}}
        \label{fig:QH_continuum}
    \end{subfigure}
    
    \begin{subfigure}[t]{0.817\textwidth}
        \includegraphics[width=\textwidth]{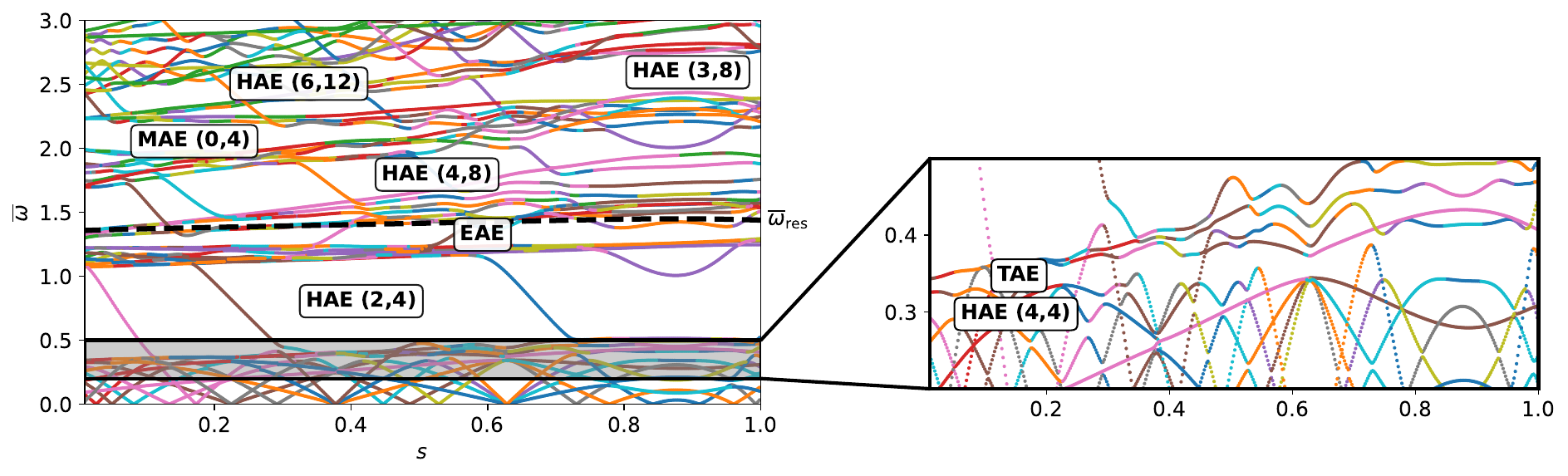}
        \caption{Ku-Boozer QH \citep{Ku_Boozer_2011}}
        \label{fig:KuBoozer_continuum}
    \end{subfigure}
    \hfill
    \begin{subfigure}[t]{0.817\textwidth}
        \includegraphics[width=\textwidth]{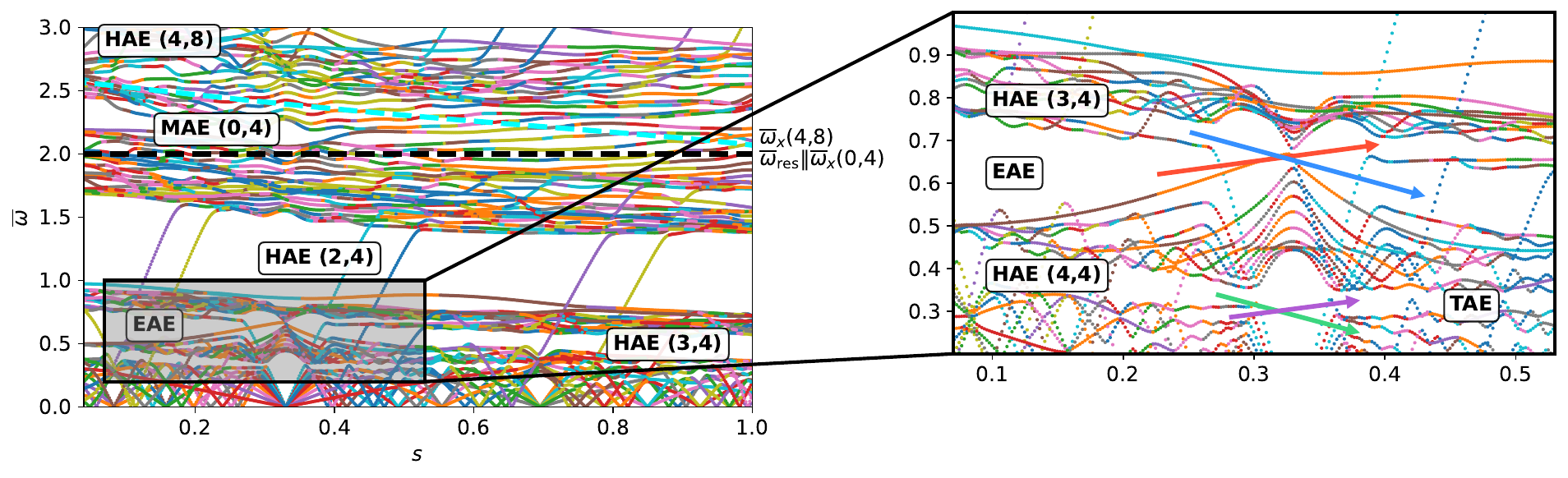}
        \caption{High-shear quasi-isodynamic (SQuID) \citep{Goodman_SQuID_2024}}
        \label{fig:QI_continuum}
    \end{subfigure}
    
    \caption{The shear Alfvén continuum for all five equilibria using the poloidal-limited mode scheme (see \figref{fig:mode_choice_cone}b) and \tabref{tab:mode_choice_params} is illustrated. The normalized frequency ($\overline{\omega}$) is $\omega / \omega_A$, as mentioned in \eqref{eq:normalized_freq}. Each color represents a distinct poloidal and toroidal Fourier mode number $m,n$. For the SQuID, $\overline{\omega}_\times(0,4)$ and $\overline{\omega}_{\mathrm{res}}$ are both at 2 and share the black dashed line. The insets show the following: $(c)$ shows the TAE and other small low-frequency gaps for the Landreman QH equilibrium, $(d)$ shows the TAE gap for the Ku-Boozer equilibrium, and $(e)$ shows the crossing of the HAE(3,4) and EAE gaps at $s \approx 0.32$ with the blue and red arrows showing the continuation of the HAE(3,4) and EAE gaps, respectively. Additionally, the inset of $(e)$ shows the crossing of the HAE(4,4) and TAE gaps with the green and purple arrows showing the continuation of the HAE(3,4) and EAE gaps, respectively.}
    \label{fig:continuum_and_gaps}
\end{figure}

We identify and label gaps according to \tabref{tab:AE_types} using the gap detection method described in \secref{sec:gap_detection}. The gap width $\Delta \overline{\omega}_{\delta m, \delta n}$ is calculated in step \gapref{c_find_avoided_crossings} as the mean width of all avoided crossings that share the same $\Delta m$, $\Delta n$. For example, the HAE(2,4) gap in \figref{fig:continuum_and_gaps}e is the average width across all $\Delta m = 2, \Delta n = 4$ avoided crossings. 

All configurations use the poloidal-limited 
mode selection scheme with the parameters in \tabref{tab:mode_choice_params}.
To ensure the mode resolution is sufficient, $m_{\text{max}}$ is chosen so that at least 10 continuum curves touch $\overline{\omega} = 0$.
As mentioned in \secref{sec:frequency_range}, we consider normalized 
frequencies $0 \le \overline{\omega} \le 3$,
which covers a range resonant with prompt fusion-born alphas, and the range is consistent with the Infinity Two stellarator study \citep{type_one_infinity_two}.

\begin{figure}
    \centering 
    \begin{subfigure}[t]{0.48\textwidth}
        \includegraphics[width=\textwidth]{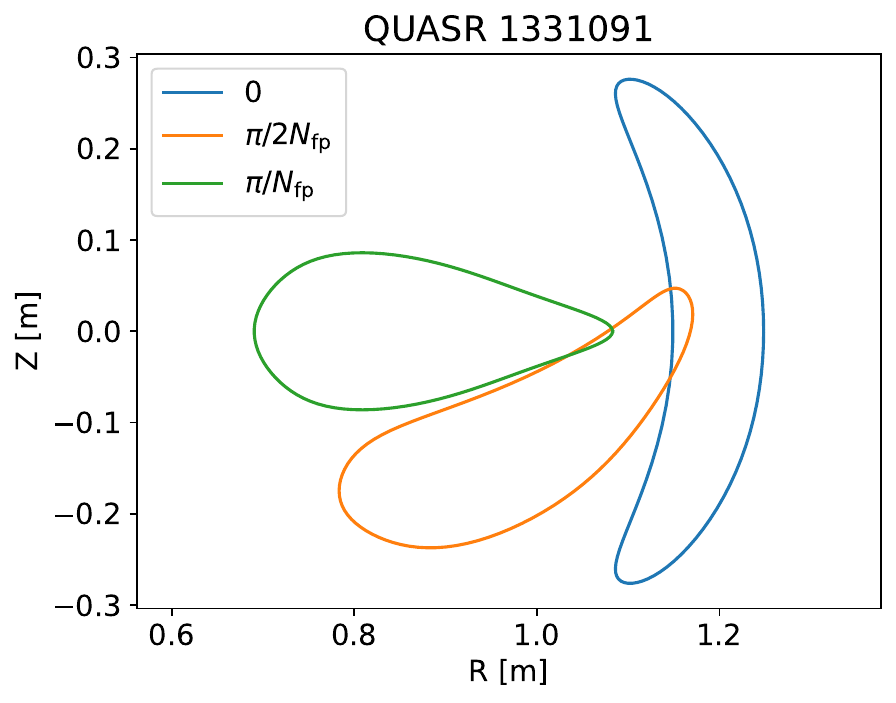}
        \caption{}
    \end{subfigure}
    \begin{subfigure}[t]{0.48\textwidth}
        \includegraphics[width=\textwidth]{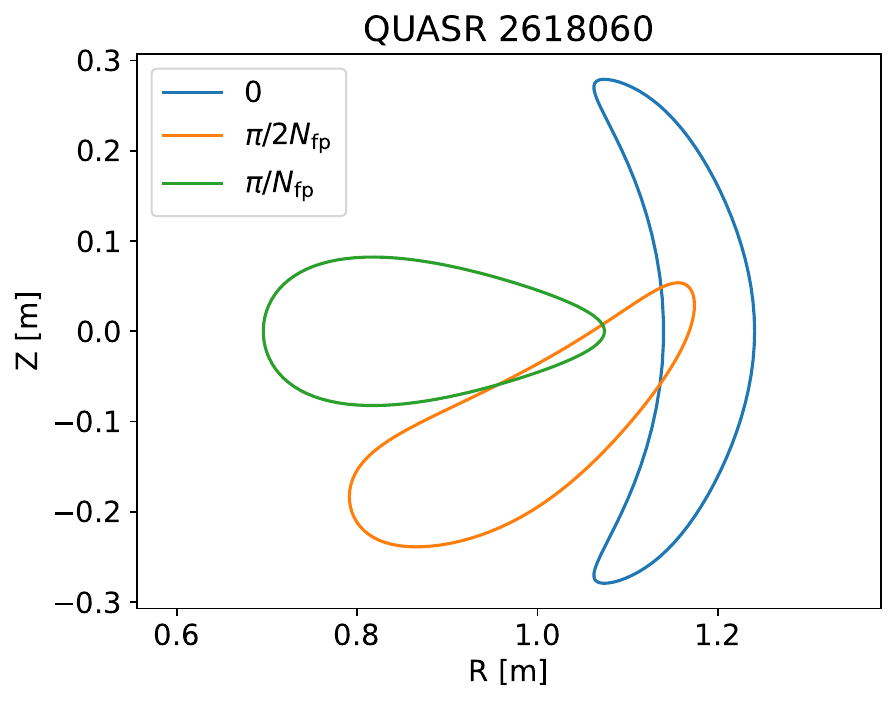}
        \caption{}
    \end{subfigure}
    \begin{subfigure}[t]{0.48\textwidth}
        \includegraphics[width=\textwidth]{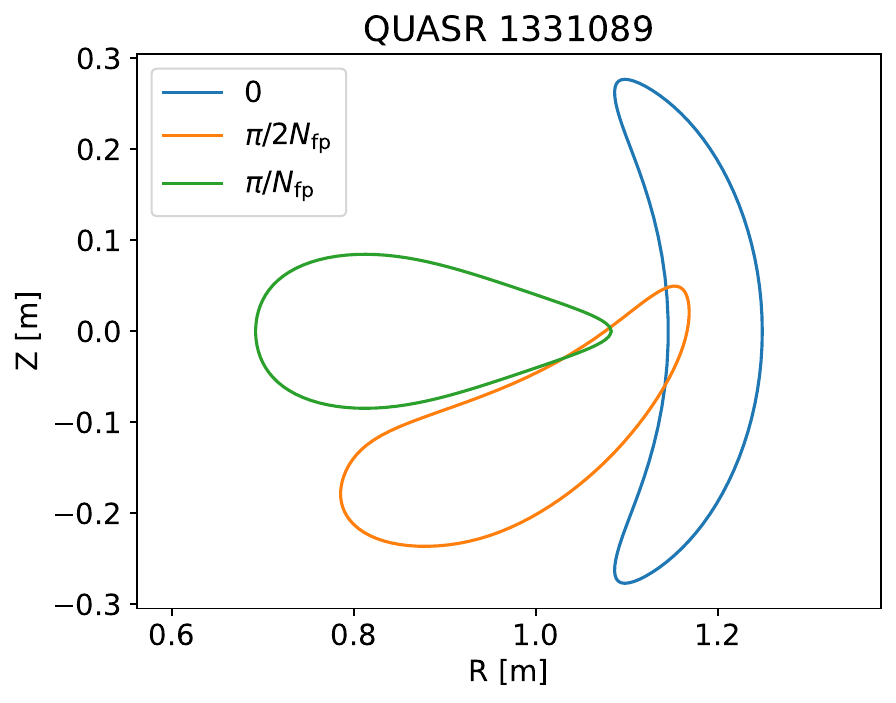}
        \caption{}
    \end{subfigure}
    \begin{subfigure}[t]{0.49\textwidth}
        \includegraphics[width=\textwidth]{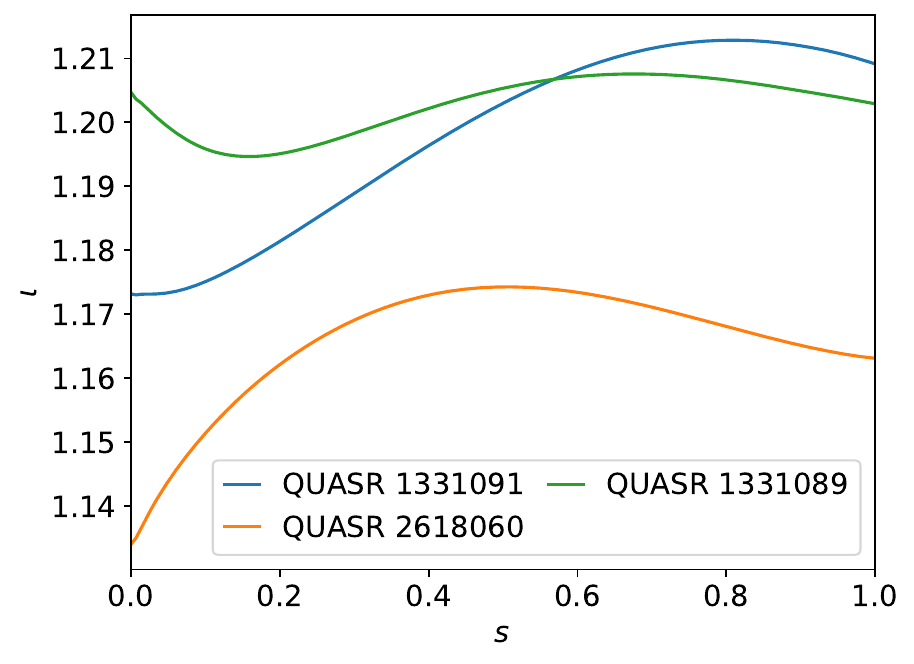}
        \caption{}
    \end{subfigure}
    \caption{Illustrated are the poloidal cross-sections at the last closed flux surface for three QUASR configurations. The configurations are in the QUASR database as 1331091, 2618060, and 1331089.}
    \label{fig:cross-sections_QUASR}
\end{figure}

\begin{figure}
    \centering 

    \includegraphics[width=\textwidth]{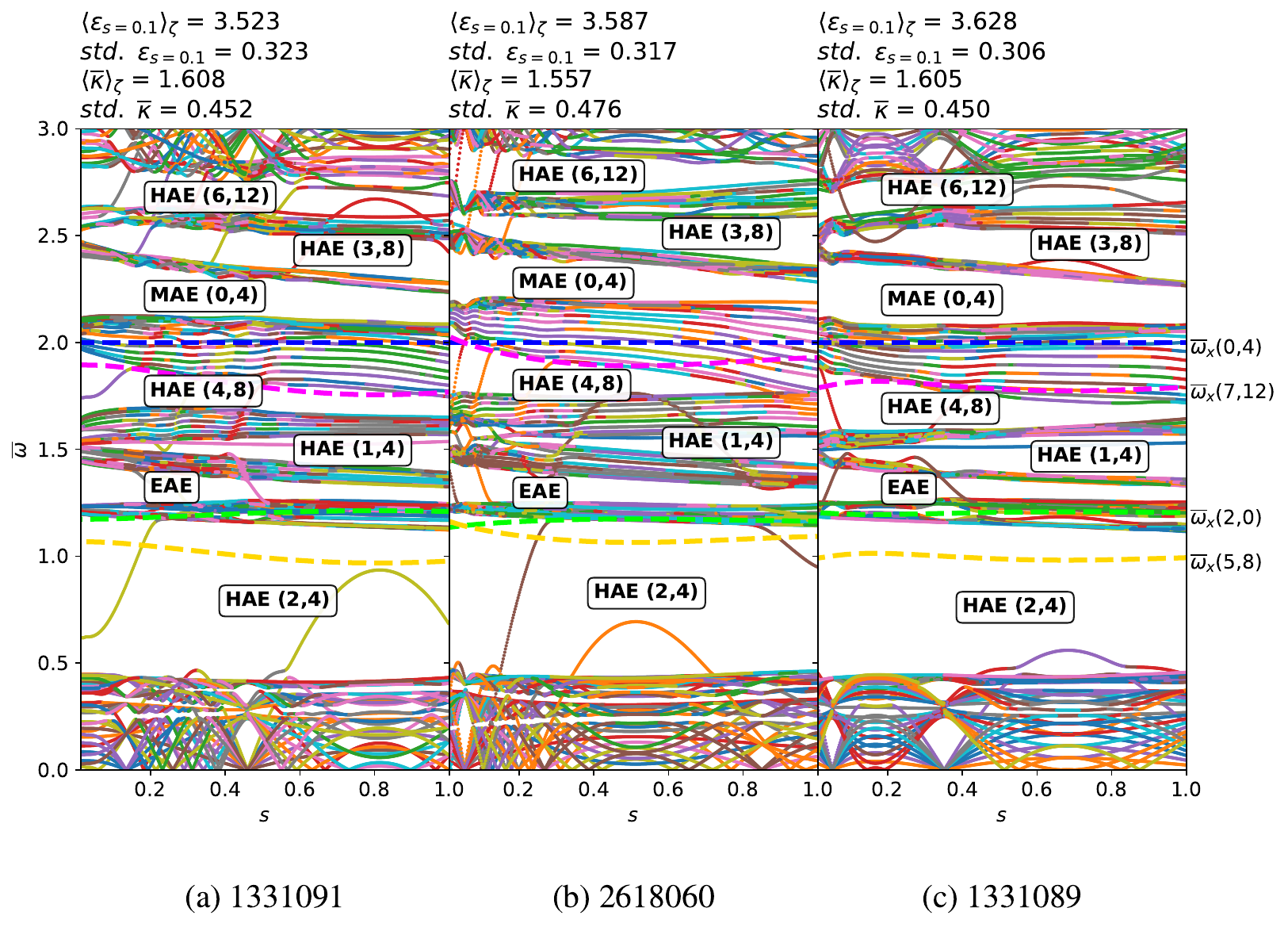}

    \caption{The continua for three QH QUASR configurations are shown \citep{Giuliani_Rodriguez_Spivak_2025}. The continua are ordered by increasing elongation and decreasing toroidal variation in elongation. The analytical gap frequency $\overline{\omega}_\times$ (see \eqref{eq:freq_crosssing}) for some particular gaps is included.}
    \label{fig:QUASR_continua}
\end{figure}

\begin{figure}
    \centering 
    \includegraphics[width=\textwidth]{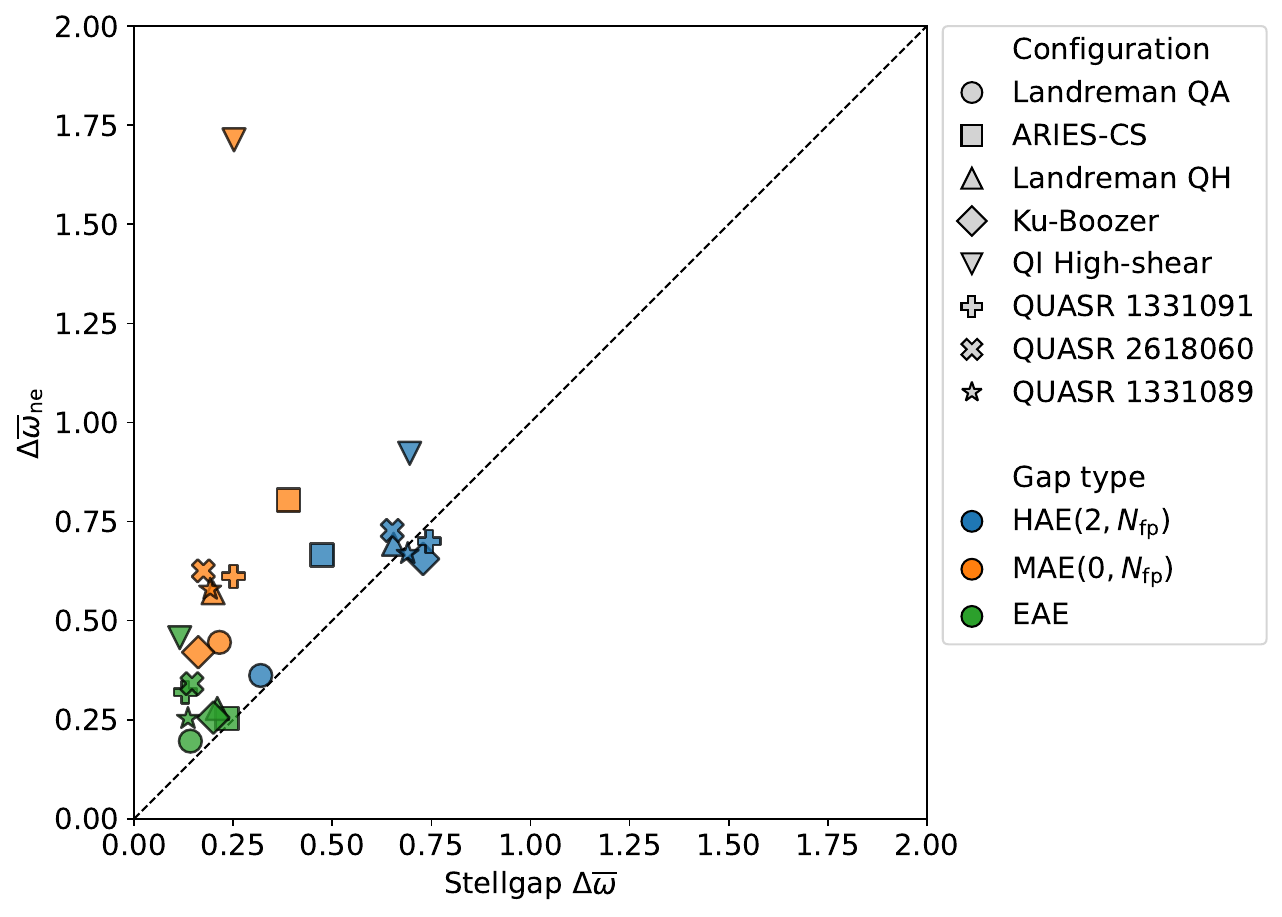}

    \caption{The Stellgap gap widths calculated using the gap detection method (see \secref{sec:gap_detection}) are plotted against the analytical gap width from near-axis theory (see \eqref{eq:analytical_width}). Both the near-axis gap width $\Delta \overline{\omega}_{\text{ne}}$ and $|\nabla \psi|^2$ are computed from the same VMEC output files used in Stellgap.}
    \label{fig:gap_widths_comparison_all}
\end{figure}

\begin{table}
    \centering
    \begin{tabular}{c||cccc}
        Configuration & field period ($N_{\text{fp}}$) & $m_{\text{max}}$ & $\iota_{\text{min}}$ & $\iota_{\text{max}}$ \\
        \hline
        Landreman $\beta=2.5\%$ QA & 2 & 20 & 0.155 & 0.510 \\
        ARIES-CS (QA) & 3 & 30 & 0.413 & 0.678 \\
        Landreman $\beta=2.5\%$ QH & 4 & 30 & 1.006 & 1.163 \\
        Ku-Boozer QH & 4 & 30 & 1.102 & 1.286 \\
        SQuID & 4 & 50 & 0.710 & 0.968 \\
    \end{tabular}
    \caption{This table contains all the parameters necessary to reproduce the Fourier modes $m,n$ used for each configuration's continuum in \figref{fig:continuum_and_gaps}. For all the configurations, $\overline{\omega}_{\max} =4$ and the $n_{\text{start}}=1$ mode family is used. The poloidal-limited mode scheme, as shown in \figref{fig:mode_choice_cone}b, is used for comparison purposes between configurations.}
    \label{tab:mode_choice_params}
\end{table}

\begin{table}
    \centering
    \begin{tabular}{c||c||c||c}
         & \makecell{Toroidal variation \\ in curvature $\kappa$} & Elongation $\varepsilon$ & \makecell{Toroidal variation \\ in elongation $\varepsilon$} \\
        \hline
        EAE & QH & QA|QH &  \\
        MAE &  &  & QA|QH \\
        HAE(2,$N_{\text{fp}}$) &  & QA|QH &  \\
        HAE(2,2$N_{\text{fp}}$) &  & QH &  \\
        Higher-order HAEs & QA|QH & QA|QH & 
    \end{tabular}
    \caption{For QA and QH, the table lists the gaps that correlate with elongation, toroidal variation in elongation, or toroidal variation in curvature according to near-axis theory \citep{Paul_Hyder_Rodriguez_Jorge_Knyazev_2025}.
    }
    \label{tab:gap_width_patterns}
\end{table}

\begin{table}
    \centering
    \begin{tabular}{c||cc||cc||cc}
        Configuration & $\left < \overline{\kappa} \right >_{\zeta}$ & std. $\overline{\kappa}$ & $\left < \varepsilon_{s=0.1} \right >_{\zeta}$ & std. $\varepsilon_{s=0.1}$ & $\left < \varepsilon_{s=1} \right >_{\zeta}$ & std. $\varepsilon_{s=1}$ \\
        \hline
        Landreman $\beta=2.5\%$ QA & 0.957 & 0.260 & 3.206 & 0.439 & 2.768 & 0.402 \\
        ARIES-CS (QA) & 0.952 & 0.464 & 2.953 & 0.401 & 1.822 & 0.211 \\
        Landreman $\beta=2.5\%$ QH & 1.691 & 0.418 & 3.187 & 0.356 & 2.452 & 0.162 \\
        Ku-Boozer QH & 1.556 & 0.455 & 4.185 & 0.235 & 2.836 & 0.032 \\
        SQuID high-shear & 1.262 & 1.007 & 3.196 & 0.636 & 1.988 & 0.161 \\
        QUASR 1331091 & 1.608 & 0.452 & 3.523 & 0.323 & 2.689 & 0.091 \\
        QUASR 2618060 & 1.557 & 0.476 & 3.590 & 0.317 & 2.745 & 0.120 \\
        QUASR 1331089 & 1.605 & 0.450 & 3.628 & 0.306 & 2.703 & 0.085 
    \end{tabular}
    \caption{
    For each configuration, the magnetic axis curvature $\kappa$ and the elongation $\varepsilon$ averaged over the toroidal angle $\zeta$ are given. The elongation is given for the 0.1 and 1.0 flux surfaces. The toroidal variation in elongation and toroidal variation in curvature are measured as the relative standard deviation $std.$}
    \label{tab:config_parameters}
\end{table}

\subsection{Gap widths for optimized equilibria without sound coupling}
\label{sec:grad_psi2_overview}

Comparing the five QA, QH, and QI configurations gives insight into the gap structure of stellarator continua. For all continua, we desire narrow gaps at EP-resonant frequencies. Wider gaps allow global modes, which avoid strong continuum damping, to exist over a broader range of frequencies. Strong resonance with fusion-born alphas at the birth energy is expected at frequencies above $\overline{\omega}_{\mathrm{res}} \equiv |M \iota-N|/2$, where $M$ is the poloidal helicity and $N$ is the toroidal helicity. The frequency threshold $\overline{\omega}_{\mathrm{res}}$ will be discussed further in \secref{sec:resonance_with_alphas}.

We now analyze the ordering of gaps by their widths. 
For all five configurations, the widest gap is the HAE(2,$\pm N_{\text{fp}}$) gap driven by rotating ellipticity \citep{Paul_Hyder_Rodriguez_Jorge_Knyazev_2025}. 
The dominant gap being an HAE is a large departure from the dominant gap in tokamaks, which is generally a TAE or EAE \citep{2014Gorelenkov}. For both QA configurations, the second widest gap is the MAE(0,$N_{\text{fp}}$) gap as illustrated in Figures \ref{fig:continuum_and_gaps}a and \ref{fig:continuum_and_gaps}b. The QH configurations have large HAE gaps, and the HAE(4,2$N_{\text{fp}}$) gap is the second widest of all the gaps, after the HAE(2,$\pm N_{\text{fp}}$) gap. Since it is a higher harmonic of the HAE(2,$\pm N_{\text{fp}}$) gap, it is also driven by rotating ellipticity \citep{Paul_Hyder_Rodriguez_Jorge_Knyazev_2025}. 
For the QH configurations, the MAE is significant but not close to the size of the widest and second-widest HAEs.

There are other distinct differences between the configurations. First, for the QA configurations, there is a lack of large gaps at high frequencies ($\overline{\omega}>2$). 
Second, in the QH configurations, the EAE is above the HAE($2, N_{\text{fp}}$) gap, while in the QA and SQuID continua, it falls below the HAE($2, N_{\text{fp}}$) gap. From \eqref{eq:freq_crosssing}, the analytical gap frequency $\overline{\omega}_\times$ for the EAE gap is $|\iota|$ and for the HAE($2, N_{\text{fp}}$) gap is $|\iota-N_{\text{fp}}/2|$. Therefore, for the EAE to be above the HAE($2, N_{\text{fp}}$) gap, $N_{\text{fp}} < 4 \iota$ must be satisfied. For the QA and QI configurations, $\iota < 1$ and $N_{\text{fp}} < 4 \iota$ is never satisfied.
Third, there is a distinctive local minimum in the Landreman QH continuum. The minimum corresponds to the flux surface where $\iota$ reverses (see \figref{fig:cross-sections}f). Such a continuum extremum can support a reverse-shear Alfvén eigenmode (RSAE) localized near that surface.

A notable feature of shear Alfvén continua is that gaps with different $\delta m, \delta n$ can interact and intersect. The crossing frequencies from \eqref{eq:freq_crosssing}, combined with the rotational transform profile, predict where such intersections occur. One of the clearest examples is in the SQuID continuum near $\overline{\omega} = 0.7$ and $s = 0.34$. As shown in the inset of \figref{fig:QI_continuum}, two pairs of gaps cross there: the EAE gap with the HAE(3,4) gap, and the TAE gap with the HAE(4,4) gap. Both pairs are predicted to intersect where $\iota = 0.8$, which the SQuID rotational transform (see \figref{fig:cross-sections}) reaches at $s \approx 0.34$. Inside this flux surface, the EAE gap lies below the HAE(3,4) gap, and outside it, the EAE gap lies above. At the intersection itself, the gaps narrow and annihilate \citep{Yakovenko_2007}, so global modes are unlikely to exist there.

Gaps can also interact without intersecting. The SQuID continuum oscillates near $\overline{\omega} = 2.5$, likely because of an interaction between the HAE(4,8) and MAE(0,4) gaps. In this region, the eigenfrequencies for a Fourier mode oscillate without an avoided crossing. These gaps are predicted to intersect where $\iota = 1$, which the SQuID equilibrium never reaches. However, the rotational transform approaches $\iota = 1$, so the analytical gap frequencies $\overline{\omega}_\times$ of the two gaps (red and blue dashed lines in \figref{fig:QI_continuum}) come close together. As a result, the gaps interact and produce an oscillation in the continuum, but they never intersect.

The relative gap widths and continuum features in quasi-symmetric configurations can be explained in part using a near-axis approach. Section 3 of \citep{Paul_Hyder_Rodriguez_Jorge_Knyazev_2025} uses near-axis theory to produce an analytical expression for the gap width. We define a coupling parameter,
\begin{align}
\epsilon = \frac{|\nabla \psi|^2}{\left< |\nabla \psi|^2 \right>} - 1,
\label{eq:epsilon}
\end{align}
where $\left< ... \right >$ is the average over the poloidal and toroidal angles.
Fourier decomposing $\epsilon$ gives
\begin{align}
 \epsilon = \sum_{\deltam,\deltan} \epsilon_{\deltam,\deltan} e^{i(\deltam \theta - \deltan \zeta)}.
 \label{eq:Fourier_epsilon}
\end{align}
Under the approximation of weak coupling, $\epsilon \ll 1$, the near-axis approximated normalized frequency shift $\Delta \overline{\omega}_{\text{ne}}$ is 
\begin{equation}
    \Delta \overline{\omega}_{\text{ne}} = 2 \overline{\omega}_{\times} |\epsilon_{\delta m,\delta n}|,
    \label{eq:analytical_width}
\end{equation}
where $\overline{\omega}_{\times}$ follows \eqref{eq:freq_crosssing} in which $\overline{\omega}_{\times}$ is the frequency of a crossing or an avoided crossing. 
The Fourier coefficients $\epsilon_{\delta m,\delta n}$ have dependencies on flux surface elongation and magnetic axis curvature $\kappa$, which can be seen in Section 3.2 of \citep{Paul_Hyder_Rodriguez_Jorge_Knyazev_2025}.
\tabref{tab:gap_width_patterns} lists the geometric drivers of each gap in QA and QH equilibria. Elongation, toroidal variation of elongation, and toroidal variation of magnetic axis curvature each drive particular gaps through their contributions to $\epsilon_{\Delta m,\Delta n}$.

From the VMEC equilibria, the elongation $\varepsilon$, toroidal variation of elongation, and toroidal variation of magnetic axis curvature $\kappa$ are obtained as follows. To estimate the elongation of different flux surfaces, we fit an ellipse to the flux surface using a least-squares method. The ellipse method works best near the magnetic axis, where concavity and triangularity of the flux surface are low. At the boundary $s=1$, the ellipse is not a great fit when shaping is strong, such as for the ARIES-CS configuration illustrated in \figref{fig:cross-sections}b. The elongation at $s=0.1$ and $s=1$ is provided in \tabref{tab:config_parameters}; however, for comparison purposes, the elongation at $s=0.1$ is used to stay near the magnetic axis and avoid strong shaping near the last closed flux surface. The toroidal variation in curvature and elongation is measured by their relative standard deviation, which is the standard deviation normalized by the mean.

Based on the values for elongation and curvature in \tabref{tab:config_parameters}, the expectations from near-axis theory can be used to explain some of the gap structure in the continua. The HAE($2, \pm N_{\text{fp}}$) gap, driven by rotating ellipticity, is the widest gap in all configurations. The MAE gap width is set by the toroidal variation of elongation, which is weaker in the QH configurations when compared to QA. Consistent with this, the MAE gap is narrower in the QH configurations than in the QA configurations. The rotational transform profile does not affect the MAE gap width according to the gap frequency \eqref{eq:freq_crosssing} and the linear scaling between gap frequency and gap width \eqref{eq:analytical_width}.

Because the gap width depends on the gap frequency \eqref{eq:analytical_width}, configurations whose widest gaps lie at lower frequencies are expected to have narrower gaps overall. At lower frequency, the gap width (see \eqref{eq:analytical_width}) is expected to be smaller. The QA continua match this expectation, since their HAE gaps lie at low frequencies ($\overline{\omega} < 2$) and are visibly narrower than in the other three configurations. Additionally, we expect the gap frequency of EAEs to be below the HAE($2, N_{\text{fp}}$) gap for the two QAs and above it for the QH and QI because of their rotational transform.

Among the four QA and QH configurations, several gaps do not follow the near-axis trends listed in \tabref{tab:gap_width_patterns}, as can be seen by comparing the geometric parameters in \tabref{tab:config_parameters} with the gap widths in \figref{fig:continuum_and_gaps}. For example, the Landreman QA has a larger elongation than both ARIES-CS and the Landreman QH, yet it has a narrower EAE gap. Similarly, ARIES-CS has a smaller elongation and magnetic axis curvature than the Landreman QA but a wider HAE($2, \pm N_{\text{fp}}$) gap, and a smaller toroidal variation of the elongation but a wider MAE gap. Part of the reason these heuristic trends break down is because they do not consider the impact of gap frequency on the gap width. Shown in \eqref{eq:analytical_width}, the gap width scales linearly with the gap frequency. For the EAE gap, the smaller rotational transform of the Landreman QA places its gap at a lower frequency, which narrows the gap. Geometric effects beyond elongation and curvature that enter through $|\nabla \psi|^2$ may also contribute to these deviations from the near-axis theory heuristic.

With the gap frequency accounted for, the near-axis theory expected gap width \ref{eq:analytical_width} correlates well with Stellgap, even when the geometric heuristic trends do not hold. As shown in \figref{fig:gap_widths_comparison_all}, the gap width from \eqref{eq:analytical_width} agrees well with Stellgap for the EAE and HAE($2, \pm N_{\text{fp}}$) gaps. The MAE gaps, however, are overestimated by the near-axis prediction. This may be due to the interaction between MAEs and adjacent HAEs, which occurs in all five configurations. All three gaps of the SQuID configuration are also overestimated, likely due to strong interactions between its gaps. Because near-axis theory assumes small coupling, it does not capture these interactions between gaps. 

The impact of the interaction between gaps is illustrated between the continua of three QUASR equilibria in \figref{fig:QUASR_continua}. QUASR is a database of quasi-symmetric stellarators \citep{Giuliani_Rodriguez_Spivak_2025}. We choose several QH equilibria with similar minor radius, aspect ratio, QS error, and mean $\iota$ to assess. As illustrated in \figref{fig:cross-sections_QUASR}, the cross-sections are similar. From (a) to (c), the elongation increases and the toroidal variation of elongation decreases. The MAE gap widths in decreasing order are (c), (a), and (b). The EAE gap widths in decreasing order are (b), (c), and (a). So, (a) and (b) follow the expected near-axis trends but (c) does not. It can be seen from the plotted $\overline{\omega}_\times$ lines that both (a) and (b) have interactions between the MAE(0,4) and HAE(7,12) gaps and interactions between the EAE and HAE(5,8) gaps. Configuration (c) lacks these interactions. The interaction between these gaps in (a) and (b) is a result of their similar $\iota$ profiles. The intersection of the MAE(0,4) and HAE(7,12) gaps and the EAE and HAE(5,8) gaps occurs when $\iota=8/7$, which is approached in (a) and achieved in (b). Configuration (c) never approaches $\iota=8/7$. Based on these and several other QUASR QH equilibria, the near-axis theory expected trends are well followed when there is a lack of interactions between gaps, and when the geometry does not strongly depart from a rotating ellipse. 

Overall, the analytically produced gap width from near-axis helps predict some aspects of the gap structure well when interactions between gaps are weak, which is dependent on the rotational transform. However, it does not hold when interactions between gaps are strong, or for strongly-shaped geometry departing from a rotating ellipse. Therefore, numerical calculations for the continua are still necessary to obtain accurate gap widths and frequencies.

\subsection{Gap widths with sound coupling}
\label{sec:grad_psi2_overview_with_sound_coupling}
To account for beta-induced Aflvén eigenmodes (BAEs), we include sound coupling under the slow sound approximation. We use deuterium ions with a density profile 
$n = n_0(1 - s^{5})$ with $n_0 = 4.8\times10^{20}\,\mathrm{m^{-3}}$ and a temperature profile of $T_e=T_0(1-s)$ with $T_0 = 11.5\,\mathrm{keV}$, as used in \citep{Bader_2021}, which is similar to the profiles used to study ARIES-CS.
A further discussion of sound coupling and its effect on continua is in \citep{2010_Konies}.

\begin{figure}
    \centering 
    \begin{subfigure}[t]{0.49\textwidth}
        \includegraphics[width=\textwidth]{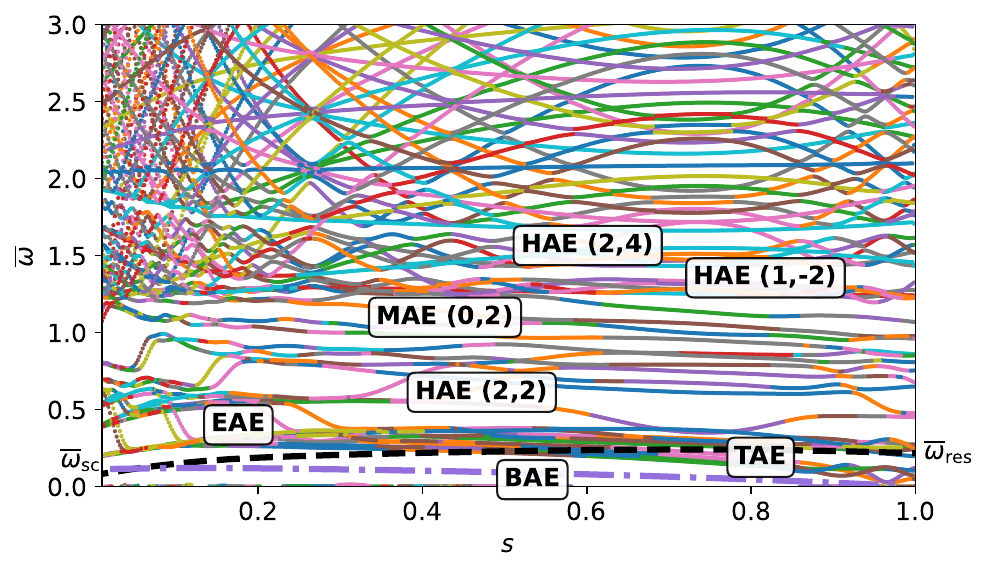}
        \caption{Landreman $\beta=2.5\%$ QA \citep{2022Landremanb}}
        \label{fig:QA_continuum_with_sound}
    \end{subfigure}
    \hfill
    \begin{subfigure}[t]{0.49\textwidth}
        \includegraphics[width=\textwidth]{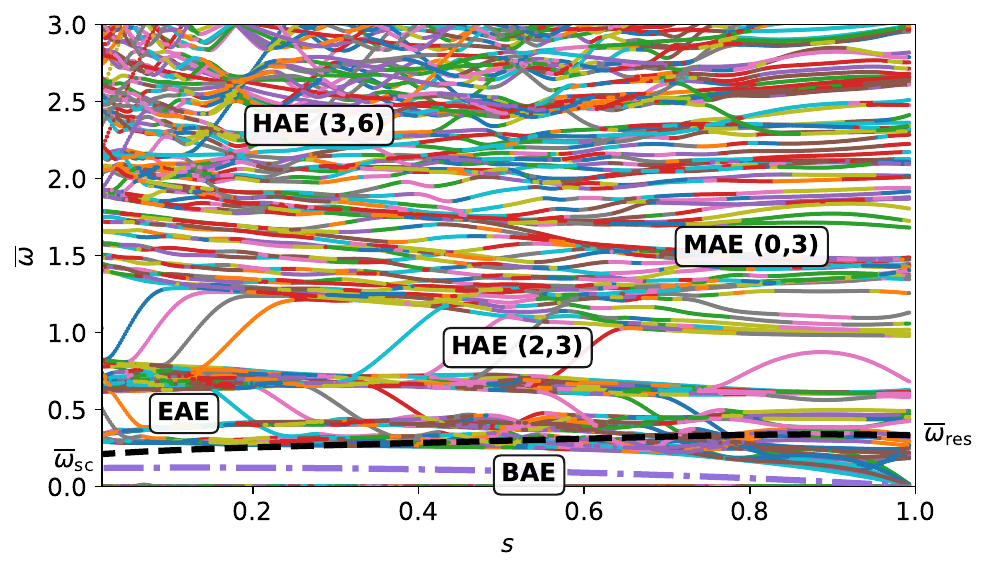}
        \caption{ARIES-CS QA \citep{Ku2008-ARIES-CS}}
        \label{fig:ARIES_continuum_with_sound}
    \end{subfigure}

    \begin{subfigure}[t]{0.49\textwidth}
        \includegraphics[width=\textwidth]{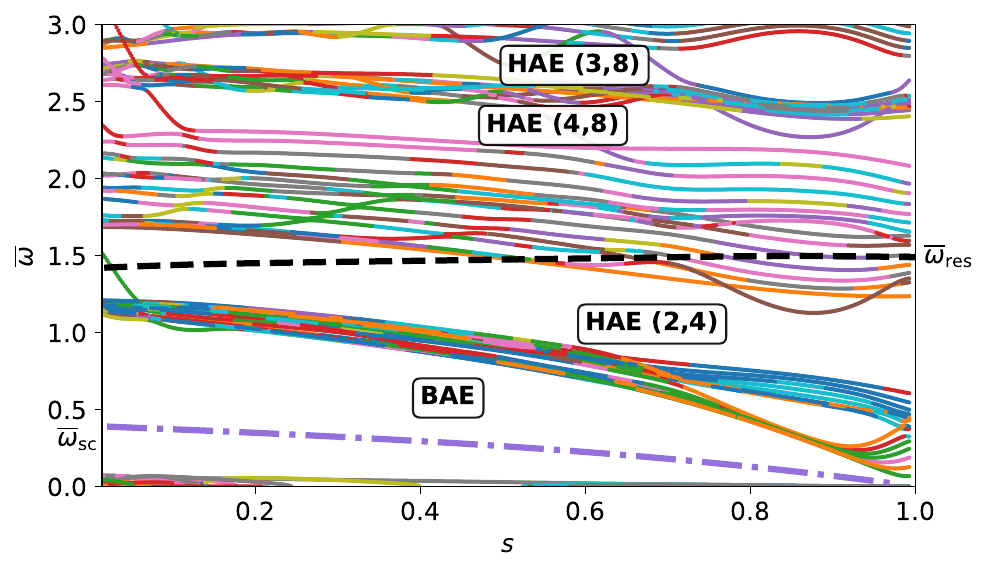}
        \caption{Landreman $\beta=2.5\%$ QH \citep{2022Landremanb} with Bader Te}
        \label{fig:QH_continuum_with_sound}
    \end{subfigure}
    \hfill
    \begin{subfigure}[t]{0.49\textwidth}
        \includegraphics[width=\textwidth]{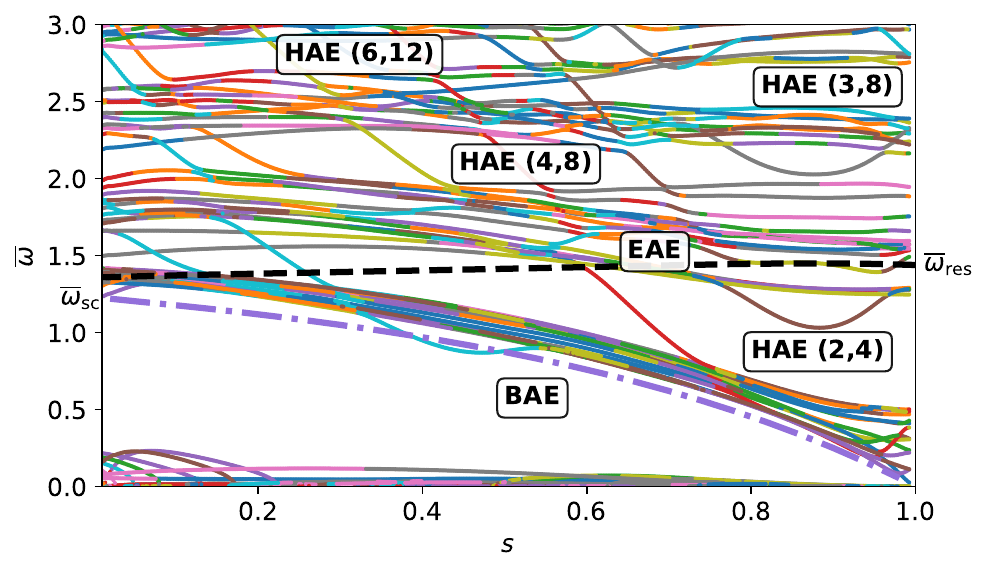}
        \caption{Ku-Boozer QH \citep{Ku_Boozer_2011} with Bader Te}
        \label{fig:KuBoozer_continuum_with_sound}
    \end{subfigure}
    
    \begin{subfigure}[t]{0.49\textwidth}
        \includegraphics[width=\textwidth]{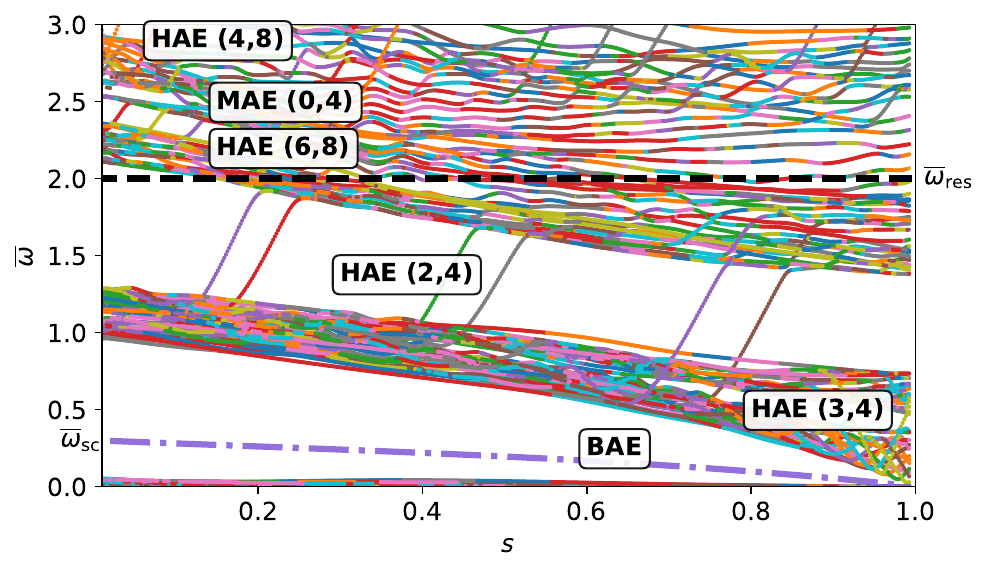}
        \caption{High-shear quasi-isodynamic (SQuID) \citep{Goodman_SQuID_2024}}
        \label{fig:QI_continuum_with_sound}
    \end{subfigure}
    
    \caption{The shear Alfvén continuum for the same five configurations as \figref{fig:continuum_and_gaps} are shown with sound coupling. The same Fourier modes are included as described in \tabref{tab:mode_choice_params}. Sound waves are included under the slow sound approximation. Shown as the purple dashed line is $\overline{\omega}_{sc}$.}
    \label{fig:continuum_and_gaps_with_sound}
\end{figure}

Including sound coupling produces a BAE gap in all five continua. 
From \citep{2010_Konies} (6)--(9), the frequency up-shift from the inclusion of sound coupling can be approximated as 
\begin{align}
\omega_{\mathrm{sc}} \equiv \sqrt{\frac{\gamma T_e}{m_i} K_{\mathrm{sc}} }
\label{eq:approx_gap_width}
\end{align}
with
\begin{align}
K_{\mathrm{sc}} \equiv \left < \kappa_g^2 \right >_{\psi} / \left < |\nabla \psi|^2 / B^2 \right >_{\psi},
\label{eq:sound_coupling_geometry}
\end{align}
where $\kappa_g = 2 (\boldsymbol{B} \times \nabla \psi) \cdot \nabla B / B^3$ is the geodesic curvature, $\left < ... \right >_{\psi} = \int d \theta d \zeta \sqrt{g} .../\int d \theta d \zeta \sqrt{g}$ is the flux-surface average, $T_e$ is the electron temperature, and $\gamma$ is the adiabatic index. The BAE gap width is driven geometrically by $K_{\mathrm{sc}}$. 
As illustrated in Figures \ref{fig:continuum_and_gaps_with_sound}c and \ref{fig:continuum_and_gaps_with_sound}e,
the analytical frequency shift from sound coupling $\overline{\omega}_{\mathrm{sc}}$ for the QI and Landreman QH equilibria underestimates the BAE gap. This may be due to additional acoustic sidebands that are not included in \eqref{eq:approx_gap_width} \citep{2010_Konies}.

\begin{figure}
    \centering 
    \includegraphics[width=\textwidth]{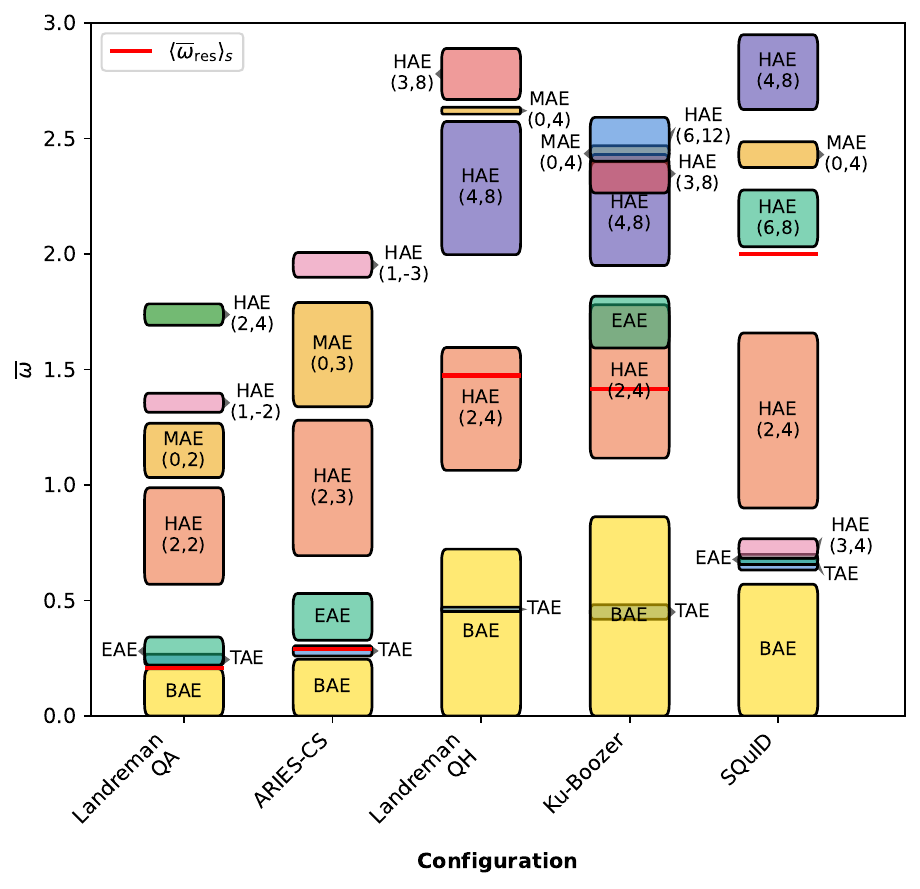}
    \caption{The gap width and frequency of select gaps are shown for all five configurations. The height of each block represents the average frequency width of the gap, and the center of the block corresponds to the average central frequency of the gap. The width is calculated using the gap detection method (\secref{sec:gap_detection}) as the average width of the avoided crossings. The fusion-born alpha resonance frequency $\left< \overline{\omega}_{\mathrm{res}} \right>_{{s}}$ is the average of $\overline{\omega}_{\mathrm{res}}$ over all flux surfaces. The BAEs are added separately from the gap detection method calculation.}
    \label{fig:gap_summation_plot}
\end{figure}

From the continua in \figref{fig:continuum_and_gaps_with_sound} and their averages in \figref{fig:gap_summation_plot}, it can be seen that the MAEs for the QH and QI are significantly reduced in gap width and the EAEs are significantly reduced. 
The change in gap width to the QA configurations is small since the BAE frequency shift is small. In general, lower-frequency gaps are reduced by a larger proportion than higher-frequency gaps due to sound coupling \citep{breizman2005plasma}. 

In these configurations, the BAE gap width is proportional to $\overline{{\omega}}_{\mathrm{sc}}$. Therefore, the BAE gap width can be minimized by reducing the geodesic curvature. However, the BAE gaps may not play an important role with fusion-born alphas in QH and QI configurations because they sit well below $\overline{\omega}_{\mathrm{res}}$, where we expect weak resonance with these alphas. In QA configurations, the BAE may be important because a portion of the gap lies above $\overline{\omega}_{\mathrm{res}}$.

\subsection{Resonance with alpha particles}
\label{sec:resonance_with_alphas}

For these reactor-scale equilibria, we are particularly interested in studying gaps with frequencies in the resonant range of fusion-born alphas at the birth energy because prompt losses are the most harmful. Gaps can contain global modes that resonate with EPs in the absence of strong continuum damping. So, small gap widths at EP resonant frequencies are desired to potentially improve EP confinement. 

We utilize a resonance condition for fusion-born alphas to determine which gaps exist at frequencies where strong resonance is likely. From Section 6 of \citep{Paul_Hyder_Rodriguez_Jorge_Knyazev_2025} and Appendix B of \citep{knyazev2026shearalfvenwaveinducedenergetic}, the resonance condition for co-passing particles is
\begin{align}
   \left \rvert \frac{\omega_A}{\omega_{\zeta}} \right \rvert = \left \rvert 1 \pm \frac{2l(M\iota - N)}{\iota \Delta m - \Delta n}\right \rvert,
   \label{eq:resonance_AE_co-passing}
\end{align}
and for counter-passing particles is
\begin{align}
    \left \rvert \frac{\omega_A}{\omega_{\zeta}} \right \rvert = \left \rvert 1 \pm \frac{4n + 2l(M \iota + N)}{\iota \Delta m - \Delta n} \right \rvert,
\end{align}
where $l$ is the sideband coupling parameter, $\omega_{\zeta}$ is the toroidal transit frequency, $N$ is the toroidal helicity, and $M$ is the poloidal helicity. In QA and QH stellarators, $M$ is 1 and in QP stellarators, it is 0. We can approximate QI stellarators as QP in order to assessing impacts of the resonance condition \citep{knyazev2026shearalfvenwaveinducedenergetic}.
We focus on co-passing alphas at the birth energy, because the co-passing resonance condition is easier to satisfy for FPPs, which often have large toroidal mode numbers $n$ \citep{2014Gorelenkov}. Using \eqref{eq:freq_crosssing} with $|l|=1$ and $\omega_A/\omega_{\zeta} < 1$ for the same reasons mentioned in \citep{Paul_Hyder_Rodriguez_Jorge_Knyazev_2025}, the resonance condition becomes
\begin{equation}
\overline{\omega} > \frac{\left|M \iota - N \right|}{2} \equiv \overline{\omega}_{\mathrm{res}}.
\label{eq:resonance_condition}
\end{equation}

The helicity $N$ of QH configurations places $\overline{\omega}_{\mathrm{res}}$ at a higher frequency than in QA configurations, so fewer of their gaps fall in the resonant range. In particular, neither the HAE(2,$N_{\text{fp}}$) gap, which is the widest gap in all configurations, nor the significant EAE gap satisfies \eqref{eq:resonance_condition}. As a result, global modes in these gaps will likely resonate only weakly with fusion-born alphas. However, the higher-frequency HAE gaps in QH and QA configurations do satisfy the condition and may contain global modes that resonate strongly with passing fusion-born alphas.

In the two QA configurations, every labeled gap in \figref{fig:continuum_and_gaps} satisfies \eqref{eq:resonance_condition} except the low-frequency BAE and TAE gaps. The resonant gaps include the three widest: the HAE(2,$N_{\text{fp}}$), MAE, and EAE gaps. The resonant HAE(2,$N_{\text{fp}}$) gap is of particular concern, because its large width allows a broad range of frequencies for alpha-resonant global modes, and avoiding it through optimization may prove difficult. This concern is partly offset by the fact that gap width scales with frequency (see \eqref{eq:analytical_width}). Because the HAE(2,$N_{\text{fp}}$) gap in QAs lies at a lower frequency, it is narrower than the same gap in QH and QI equilibria. Even so, QH and QI configurations may be easier to optimize for alpha confinement, because their largest gaps lie outside the region of strong resonance.

Additional analysis of specific global modes would be required to determine whether there is resonance within the frequency region and how strong the resonance is. Because many of their gaps have frequencies that satisfy \eqref{eq:resonance_condition}, the QA configurations are likely to support global modes that resonate with fusion-born alpha particles. Further analysis is needed to determine the strength of this resonance, and nonlinear analysis can reveal where the mode would saturate.

Although prompt fusion-born alphas are of greatest concern, a slowing-down distribution develops over time. 
These lower-energy alphas resonate at lower frequencies, so they may interact with global modes below $\overline{\omega}_{\mathrm{res}}$, including modes in the largest QH and QI gaps such as the HAE(2,$N_{\text{fp}}$) gap. Gaps below this threshold may therefore still be a concern, both for slowed-down alphas and for other EPs such as neutral-beam ions.

\FloatBarrier

\section{Conclusions}
\label{conclusion}

In this paper, some of the numerical considerations when calculating the shear Alfvén continuum are discussed, including mode selection in \secref{sec:numerics}. The three mode choice schemes presented, poloidal-limited, toroidal-limited, and block, all give similar continua as illustrated in \figref{fig:gap_width_difference}. As a result, the poloidal-limited mode scheme is used for comparison purposes and for calculating the continua in Figures \ref{fig:continuum_and_gaps} and \ref{fig:continuum_and_gaps_with_sound} and the gaps in \figref{fig:gap_widths_comparison_all}.

A method for identifying and labeling shear Alfvén continuum gaps is discussed in \secref{sec:gap_detection}. In addition to labeling the gaps, the method also produces the width and frequency for individual avoided crossings and gaps. We use this to compare against the near-axis expected gap width \eqref{eq:analytical_width} in \figref{fig:gap_widths_comparison_all}. The near-axis gap width $\Delta \overline{\omega}_{\text{ne}}$ matches well for the HAE(2,$N_{\text{fp}}$), where $N_{\text{fp}}$ is the field period, and EAE gaps in the QA and QH configurations, but agreement is poor for the MAE gaps and the SQuID configuration.

Near-axis theory reveals which geometric properties, such as elongation or magnetic axis curvature, affect each gap, as shown in \tabref{tab:gap_width_patterns}. However, near-axis expectations and predicted gap widths may fail when the interaction between gaps is strong, which likely explains the poor agreement for the MAE gaps and the SQuID configuration. The interaction and intersection of gaps depend on the rotational transform.

A comparison of the continua between different QA, QH, and QI configurations reveals some similarities in the continuum gap structure of the equilibria. Driven by rotating ellipticity, the HAE(2,$N_{\text{fp}}$) gap is the dominant gap for all five configurations across QA, QH, and QI. We expect strong resonance with prompt-lost passing fusion-born alpha particles in gaps with frequencies above $\overline{\omega}_{\mathrm{res}} \equiv \left|M \iota - N \right| / 2$, where $\iota$ is the rotational transform, $M$ is the poloidal helicity, and $N$ is the toroidal helicity (see \eqref{eq:resonance_condition}). Global modes in these gaps will therefore likely resonate strongly with fusion-born alphas, whereas those in gaps below $\overline{\omega}_{\mathrm{res}}$ should resonate only weakly. Because $N=0$ in the QAs, $\overline{\omega}_{\mathrm{res}}$ falls at low frequency, below the HAE($2, N_{\text{fp}}$) gap. \secref{sec:resonance_with_alphas} shows that, for the QA equilibria, all gaps in \figref{fig:continuum_and_gaps} except the low-frequency TAE gaps are expected to resonate strongly with these alphas. The TAE for all the configurations is small and below $\overline{\omega}_{\mathrm{res}}$. The QH equilibria are not expected to have strong resonance for the HAE(2,$N_{\text{fp}}$) and EAE gaps. This is significant since the HAE(2,$N_{\text{fp}}$) gap has the largest gap width. The inclusion of slow sound coupling introduces a frequency shift in the form of a BAE. The frequency shift follows \eqref{eq:approx_gap_width}, indicating that larger geodesic curvature increases the frequency of the BAE. In QAs, the BAE gaps sit around $\overline{\omega}_{\mathrm{res}}$, so they may have strong resonance with fusion-born alphas. Although, BAEs may not be a concern for fusion-born alphas in QH and QI because the BAEs sit below $\overline{\omega}_{\mathrm{res}}$.

In the future, the equilibria can be optimized further to reduce their gap widths at frequencies resonant with energetic particles. However, because all QA gaps except the TAE gaps are resonant with fusion-born alphas, it may prove difficult to optimize QA gaps away from resonance with alpha particles. Optimization using near-axis theory is performed in \citep{Paul_Hyder_Rodriguez_Jorge_Knyazev_2025}. However, as seen in \secref{sec:continuum_gaps}, expectations from near-axis theory can fail when interactions between gaps are strong.

\section*{Acknowledgements}

The authors would like to acknowledge discussions with D. Spong and J. Varela. Additionally, the authors would like to acknowledge discussions with J. Larson in developing the gap detection method. This work was supported by the National Science Foundation under award 2512171 and by the U.S. Department of Energy under contracts DE-AC02-09CH11466 (StellFoundry), DE-SC0024548 (HiFiStell), and DE-SC0024630. This research used resources of the National Energy Research Scientific Computing Center (NERSC), a Department of Energy Office of Science User Facility using NERSC award ERCAP0031926. The authors used Claude Code (Anthropic) between August and November 2026 to assist with grammar and sentence structure in the manuscript and to clean up and modify the analysis scripts.

\appendix
\section{Further details on the gap detection method}
\label{app:appendix_a}

\subsubsection{\textbf{Step} \gapref{c_find_avoided_crossings}}
\label{app:Step_C}

From step \gapref{b_finding_switches}, we get the Fourier mode switches along a continuum curve. In order to find if there is an avoided crossing, we look for pairs of these switches. We traverse all the found switches and identify those with the same mode numbers in the opposite order. For example, a switch from $m_1,n_1$ to $m_2,n_2$ on one continuum curve and a switch from $m_2,n_2$ to $m_1,n_1$ on another match. The match that is closest in $s$ forms a matching pair, with ties broken by choosing the one closest in frequency. 
A matching pair is labeled as an avoided crossing only if their separation in $s$ is at most $\Delta s_{\text{max}}$. In this paper, we use $\Delta s_{\text{max}} = 0.1$.

The frequency gap between the two switch locations determines the width $\Delta \overline{\omega}_{\Delta m, \Delta n}$ of the avoided crossing. The location of the avoided crossing is the average between the radial $s$ and frequency $\overline{\omega}$ locations of the two mode switches. After this step, we have $\Delta m$ and $\Delta n$ from the difference between $m_1,n_1$ and $m_2,n_2$, the avoided crossing's width $\Delta \overline{\omega}_{\Delta m, \Delta n}$, and the radial position $s$ and frequency $\overline{\omega}$ of the avoided crossing.

\subsubsection{\textbf{Step} \gapref{d_gap_labeling}}
\label{app:Step_D}

The avoided crossings are grouped by their mode label ($\Delta m, \Delta n$). These groups of avoided crossings can be labeled as gaps after a couple of filters.

The first filter accounts for some irregular avoided crossings. An example of an irregular gap is one in which several avoided crossings of the same $\Delta m, \Delta n$ fall on one flux surface. Another example is one in which all the avoided crossings for a gap with mode numbers ($\Delta m,\Delta n$) are far apart, in terms of frequency, such that no continuous gap is discernible. Two filters are used to remove irregular avoided crossings. The latter filter, which is mentioned in \secref{sec:gap_detection}, checks that the number of avoided crossings is at least $d$.

The first filter uses the maximum width of all avoided crossings with the same $\Delta m,\Delta n$: $\Delta \overline{\omega}_{\Delta m,\Delta n, \text{max}}$. The filter removes all avoided crossings whose difference in frequency $\overline{\omega}$ from the next closest avoided crossing with the same $\Delta m,\Delta n$ (i.e., belonging to the same gap $\delta m,\delta n$) is greater than $\Delta \overline{\omega}_{\Delta m,\Delta n, \text{max}}$, ignoring radial position $s$.
An avoided crossing with a frequency greater than $\Delta \overline{\omega}_{\Delta m,\Delta n, \text{max}}$ from the next closest avoided crossing is well outside the gap when using a normalized continuum. 

The first step in implementing this filter is to check each gap $\delta m$, $\delta n$ as a whole.
We find $\Delta \overline{\omega}_{\Delta m,\Delta n, \text{max}}$ and then check if any two avoided crossings with $\Delta m=\delta m,\Delta n=\delta n$ are within $\Delta \overline{\omega}_{\Delta m,\Delta n, \text{max}}$.
If not, that gap is not labeled, since it either lacks enough modes to be labeled correctly, in which case the mode resolution should be increased, or the avoided crossings do not identify a gap. 
So, we check that the maximum distance in frequency for one crossing with at least one other avoided crossing of that gap $\delta m$, $\delta n$ is less than the maximum gap width $\Delta \overline{\omega}_{\text{max}}$ for that gap. In the \figref{fig:continuum_and_gaps}c QH $\beta=2.5\%$, for example, an avoided crossing with a normalized frequency of $\overline{\omega} = 2.5$ and $\delta m = 4, \delta n = 2$ should not be included in the HAE(2,4) gap, which has an average normalized frequency of about 0.9.

Effectively, with the first filter, any standalone $\Delta m,\Delta n$ avoided crossings are not labeled, where it stands alone because the avoided crossing is too far in frequency space from every other avoided crossing belonging to the $\delta m,\delta n$ gap.

After the first filter removes some irregular avoided crossings, the second filter checks that the number of avoided crossings identifying a particular gap exceeds the minimum $d$, typically 3. If it does, a gap label $\delta m$, $\delta n$ with the corresponding name from \tabref{tab:AE_types} is added at the average frequency of the avoided crossings. If it does not, that gap is not labeled. If the user needs that gap to be identified and the number of crossings less than $d$, the resolution of the mode scheme should be increased, as discussed in \secref{sec:mode_schemes}. This can be done by increasing $m_{\text{max}}$ or $n_{\text{max}}$ in the poloidal-limited or toroidal-limited mode schemes, respectively.

\bibliographystyle{jpp}

\bibliography{jpp-instructions}

@article{Paul_Hyder_Rodriguez_Jorge_Knyazev_2025, title={The shear Alfvén continuum of quasisymmetric stellarators}, volume={91}, DOI={10.1017/S0022377825100524}, number={4}, journal={Journal of Plasma Physics}, author={Paul, Elizabeth J. and Hyder, Abdullah and Rodríguez, Eduardo and Jorge, Rogério and Knyazev, Alexey}, year={2025}, pages={E101}}

@article{breizman2005plasma,
  title={Plasma pressure effect on Alfv{\'e}n cascade eigenmodes},
  author={Breizman, BN and Pekker, MS and Sharapov, SE and others},
  journal={Physics of Plasmas},
  volume={12},
  number={11},
  year={2005},
  publisher={AIP Publishing}
}

@article{Bader_2021,
doi = {10.1088/1741-4326/ac2991},
url = {https://doi.org/10.1088/1741-4326/ac2991},
year = {2021},
month = {oct},
publisher = {IOP Publishing},
volume = {61},
number = {11},
pages = {116060},
author = {Bader, A. and Anderson, D.T. and Drevlak, M. and Faber, B.J. and Hegna, C.C. and Henneberg, S. and Landreman, M. and Schmitt, J.C. and Suzuki, Y. and Ware, A.},
title = {Modeling of energetic particle transport in optimized stellarators},
journal = {Nuclear Fusion}
}

@article{Greenbaum_2020, title={First-Order Perturbation Theory for Eigenvalues and Eigenvectors}, volume={62}, ISSN={1095-7200}, url={http://dx.doi.org/10.1137/19M124784X}, DOI={10.1137/19m124784x}, number={2}, journal={SIAM Review}, publisher={Society for Industrial & Applied Mathematics (SIAM)}, author={Greenbaum, Anne and Li, Ren-Cang and Overton, Michael L.}, year={2020}, month=Jan, pages={463–482} }

@techreport{Giles2008,
  author      = {Giles, Michael B.},
  title       = {An Extended Collection of Matrix Derivative Results for Forward and Reverse Mode Algorithmic Differentiation},
  institution = {Oxford University Computing Laboratory},
  address     = {Oxford, UK},
  type        = {Technical Report},
  number      = {NA-08/01},
  year        = {2008},
  month       = jan
}

@article{Murthy1988,
  author  = {Murthy, Durbha V. and Haftka, Raphael T.},
  title   = {Derivatives of Eigenvalues and Eigenvectors of a General Complex Matrix},
  journal = {International Journal for Numerical Methods in Engineering},
  volume  = {26},
  number  = {2},
  pages   = {293--311},
  year    = {1988},
  doi     = {10.1002/nme.1620260202}
}

@article{knyazev2026shearalfvenwaveinducedenergetic, title={On shear-Alfvén wave-induced energetic ion transport in optimised stellarators}, volume={92}, DOI={10.1017/S0022377826101974}, number={4}, journal={Journal of Plasma Physics}, author={Knyazev, Alexey Romanovich and Lachmann, Alexandra L. and Goodman, Alan Geoffrey and Hyder, Abdullah and Czekanski, Michael and Spong, Donald and Paul, Elizabeth J.}, year={2026}, pages={E104}}

@article{Kruger_1998,
    author = {Kruger, S. E. and Hegna, C. C. and Callen, J. D.},
    title = {Generalized reduced magnetohydrodynamic equations},
    journal = {Physics of Plasmas},
    volume = {5},
    number = {12},
    pages = {4169-4182},
    year = {1998},
    month = {12},
    issn = {1070-664X},
    doi = {10.1063/1.873152},
    url = {https://doi.org/10.1063/1.873152},
    eprint = {https://pubs.aip.org/aip/pop/article-pdf/5/12/4169/19225653/4169_1_online.pdf},
}

@article{type_one_infinity_two, title={Alpha-particle confinement in Infinity Two Fusion Pilot Plant baseline plasma design}, volume={91}, DOI={10.1017/S0022377825000352}, number={4}, journal={Journal of Plasma Physics}, author={Carbajal, L. and Varela, J. and Bader, A. and Guttenfelder, W. and Cerfon, A. and Schmitt, J.C. and Morrissey, J. and Hegna, C.C. and Canik, J.M. and Mandell, N.R. and et al.}, year={2025}, pages={E94}}

@article{Ku_Boozer_2011,
doi = {10.1088/0029-5515/51/1/013004},
url = {https://doi.org/10.1088/0029-5515/51/1/013004},
year = {2010},
month = {dec},
publisher = {},
volume = {51},
number = {1},
pages = {013004},
author = {Ku, L.P. and Boozer, A.H.},
title = {New classes of quasi-helically symmetric stellarators},
journal = {Nuclear Fusion}
}

@article{Landreman_paul_preciseQA,
  title = {Magnetic Fields with Precise Quasisymmetry for Plasma Confinement},
  author = {Landreman, Matt and Paul, Elizabeth},
  journal = {Phys. Rev. Lett.},
  volume = {128},
  issue = {3},
  pages = {035001},
  numpages = {6},
  year = {2022},
  month = {Jan},
  publisher = {American Physical Society},
  doi = {10.1103/PhysRevLett.128.035001},
  url = {https://link.aps.org/doi/10.1103/PhysRevLett.128.035001}
}

@article{Liu_Chen_1995,
doi = {10.1088/0031-8949/1995/T60/011},
url = {https://doi.org/10.1088/0031-8949/1995/T60/011},
year = {1995},
month = {jan},
publisher = {},
volume = {1995},
number = {T60},
pages = {81},
author = {Liu Chen and Fulvio Zonca},
title = {Theory of shear Alfvén waves in toroidal plasmas},
journal = {Physica Scripta}
}

@article{Goodman_SQuID_2024,
  title = {Quasi-Isodynamic Stellarators with Low Turbulence as Fusion Reactor Candidates},
  author = {Goodman, Alan G. and Xanthopoulos, Pavlos and Plunk, Gabriel G. and Smith, H\aa{}kan and N\"uhrenberg, Carolin and Beidler, Craig D. and Henneberg, Sophia A. and Roberg-Clark, Gareth and Drevlak, Michael and Helander, Per},
  journal = {PRX Energy},
  volume = {3},
  issue = {2},
  pages = {023010},
  numpages = {11},
  year = {2024},
  month = {Jun},
  publisher = {American Physical Society},
  doi = {10.1103/PRXEnergy.3.023010},
  url = {https://link.aps.org/doi/10.1103/PRXEnergy.3.023010}
}

@article{salat_shear_2001,
	title = {Shear {Alfvén} mode resonances in nonaxisymmetric toroidal low-pressure plasmas. {I}. {Mode} equations in arbitrary geometry},
	volume = {8},
	issn = {1070-664X, 1089-7674},
	url = {https://pubs.aip.org/pop/article/8/4/1200/264748/Shear-Alfven-mode-resonances-in-nonaxisymmetric},
	doi = {10.1063/1.1352057},
	language = {en},
	number = {4},
	urldate = {2024-05-17},
	journal = {Physics of Plasmas},
	author = {Salat, A. and Tataronis, J. A.},
	month = apr,
	year = {2001},
	pages = {1200--1206},
}

@article{Ku2008-ARIES-CS,
author = {L. P. Ku and P. R. Garabedian and J. Lyon and A. Turnbull and A. Grossman and T. K. Mau and M. Zarnstorff and ARIES team},
title = {Physics Design for ARIES-CS},
journal = {Fusion Science and Technology},
volume = {54},
number = {3},
pages = {673--693},
year = {2008},
publisher = {American Nuclear Society},
doi = {10.13182/FST08-A1899},
URL = {https://doi.org/10.13182/FST08-A1899},
eprint = {https://doi.org/10.13182/FST08-A1899}
}

@article{Rosenbluth_and_Rutherford_1975,
  title = {Excitation of Alfv\'en Waves by High-Energy Ions in a Tokamak},
  author = {Rosenbluth, M. N. and Rutherford, P. H.},
  journal = {Phys. Rev. Lett.},
  volume = {34},
  issue = {23},
  pages = {1428--1431},
  numpages = {0},
  year = {1975},
  month = {Jun},
  publisher = {American Physical Society},
  doi = {10.1103/PhysRevLett.34.1428},
  url = {https://link.aps.org/doi/10.1103/PhysRevLett.34.1428}
}

@inproceedings{
cadena2025constellaration,
title={ConStellaration: A dataset of {QI}-like stellarator plasma boundaries and optimization benchmarks},
author={Santiago A Cadena and Andrea Merlo and Emanuel Laude and Alexander Bauer and Atul Agrawal and Maria Pascu and Marija Savtchouk and Lukas Bonauer and Enrico Guiraud and Stuart R. Hudson and Markus Kaiser},
booktitle={The Thirty-ninth Annual Conference on Neural Information Processing Systems Datasets and Benchmarks Track},
year={2025},
url={https://openreview.net/forum?id=NQSbGKlCpx}
}

@article{Schwab_1993,
    author = {Schwab, Carolin},
    title = {Ideal magnetohydrodynamics: Global mode analysis of three‐dimensional plasma configurations},
    journal = {Physics of Fluids B: Plasma Physics},
    volume = {5},
    number = {9},
    pages = {3195-3206},
    year = {1993},
    month = {09},
    issn = {0899-8221},
    doi = {10.1063/1.860656},
    url = {https://doi.org/10.1063/1.860656},
    eprint = {https://pubs.aip.org/aip/pfb/article-pdf/5/9/3195/12355794/3195_1_online.pdf},
}

@article{Konies_and_Kleiber,
    author = {Könies, Axel and Kleiber, Ralf},
    title = {A computational approach to continuum damping of Alfvén waves in two and three-dimensional geometry},
    journal = {Physics of Plasmas},
    volume = {19},
    number = {12},
    pages = {122111},
    year = {2012},
    month = {12},
    issn = {1070-664X},
    doi = {10.1063/1.4769115},
    url = {https://doi.org/10.1063/1.4769115},
    eprint = {https://pubs.aip.org/aip/pop/article-pdf/doi/10.1063/1.4769115/14081476/122111_1_online.pdf},
}

@article{Bader_Wistell-A, title={Advancing the physics basis for quasi-helically symmetric stellarators}, volume={86}, DOI={10.1017/S0022377820000963}, number={5}, journal={Journal of Plasma Physics}, author={Bader, A. and Faber, B. J. and Schmitt, J. C. and Anderson, D. T. and Drevlak, M. and Duff, J. M. and Frerichs, H. and Hegna, C. C. and Kruger, T. G. and Landreman, M. and et al.}, year={2020}, pages={905860506}}

@article{Riemann_2025,
  title = {Excitation of Alfv\'enic Modes via Electromagnetic Turbulence in Wendelstein 7-X},
  author = {Riemann, J\"org and Mendes, Sara Vaz and Rahbarnia, Kian and Kleiber, Ralf and Slaby, Christoph and K\"onies, Axel and Borchardt, Matthias and Mishchenko, Alexey and Thomsen, Henning and B\"uschel, Charlotte and von Stechow, Adrian and the Wendelstein 7-X Team and B\"ahner, Jan-Peter and Hansen, S\o{}ren Kjer and Edlund, Eric},
  journal = {Phys. Rev. Lett.},
  volume = {134},
  issue = {2},
  pages = {025103},
  numpages = {6},
  year = {2025},
  month = {Jan},
  publisher = {American Physical Society},
  doi = {10.1103/PhysRevLett.134.025103},
  url = {https://link.aps.org/doi/10.1103/PhysRevLett.134.025103}
}

@article{Zonca_1999,
    author = {Zonca, Fulvio and Chen, Liu and Dong, J. Q. and Santoro, Robert A.},
    title = {Existence of ion temperature gradient driven shear Alfvén instabilities in tokamaks},
    journal = {Physics of Plasmas},
    volume = {6},
    number = {5},
    pages = {1917-1924},
    year = {1999},
    month = {05},
    issn = {1070-664X},
    doi = {10.1063/1.873449},
    url = {https://doi.org/10.1063/1.873449},
    eprint = {https://pubs.aip.org/aip/pop/article-pdf/6/5/1917/19072859/1917_1_online.pdf},
}

@article{Heidbrink_2021,
doi = {10.1088/1741-4326/abc4c3},
url = {https://doi.org/10.1088/1741-4326/abc4c3},
year = {2020},
month = {dec},
publisher = {IOP Publishing},
volume = {61},
number = {1},
pages = {016029},
author = {Heidbrink, W.W. and Van Zeeland, M.A. and Austin, M.E. and Bierwage, A. and Chen, Liu and Choi, G.J. and Lauber, P. and Lin, Z. and McKee, G.R. and Spong, D.A.},
title = {‘BAAE’ instabilities observed without fast ion drive},
journal = {Nuclear Fusion}
}

@article{Stellgap,
  title={Shear {Alfv{\'e}n} continua in stellarators},
  author={Spong, D. A. and Sanchez, R. and Weller, A.},
  journal={Physics of Plasmas},
  volume={10},
  number={8},
  pages={3217--3224},
  year={2003},
  publisher={American Institute of Physics}
}

@article{1991Garren,
  title={Magnetic field strength of toroidal plasma equilibria},
  author={Garren, D. A. and Boozer, A. H.},
  journal={Physics of Fluids B: Plasma Physics},
  volume={3},
  number={10},
  pages={2805--2821},
  year={1991},
  publisher={American Institute of Physics}
}

@article{2022Landremanb,
  title={Optimization of quasi-symmetric stellarators with self-consistent bootstrap current and energetic particle confinement},
  author={Landreman, M. and Buller, S. and Drevlak, M.},
  journal={Physics of Plasmas},
  volume={29},
  pages={082501},
  number={8},
  doi = {10.1063/5.0098166},
  year={2022}
}

@article{2010_Konies,
    author = {Könies, Axel and Eremin, Denis},
    title = {Coupling of Alfvén and sound waves in stellarator plasmas},
    journal = {Physics of Plasmas},
    volume = {17},
    number = {1},
    pages = {012107},
    year = {2010},
    month = {01},
    issn = {1070-664X},
    doi = {10.1063/1.3274921},
    url = {https://doi.org/10.1063/1.3274921},
    eprint = {https://pubs.aip.org/aip/pop/article-pdf/doi/10.1063/1.3274921/15667254/012107_1_online.pdf},
}

@article{Yakovenko_2007,
doi = {10.1088/0741-3335/49/4/015},
url = {https://doi.org/10.1088/0741-3335/49/4/015},
year = {2007},
month = {mar},
publisher = {},
volume = {49},
number = {4},
pages = {535},
author = {Yakovenko, Yu V and Weller, A and Werner, A and Zegenhagen, S and Fesenyuk, O P and Kolesnichenko, Ya I},
title = {Poloidal trapping of the high-frequency Alfvén continuum and eigenmodes in stellarators},
journal = {Plasma Physics and Controlled Fusion}
}

@article{Giuliani_Rodriguez_Spivak_2025, title={A comprehensive exploration of quasisymmetric stellarators and their coil sets}, volume={91}, DOI={10.1017/S0022377825000509}, number={5}, journal={Journal of Plasma Physics}, author={Giuliani, Andrew and Rodríguez, Eduardo and Spivak, Marina}, year={2025}, pages={E128}}

@article{Swanson_2025,
doi = {10.1088/1741-4326/ada56a},
url = {https://doi.org/10.1088/1741-4326/ada56a},
year = {2025},
month = {jan},
publisher = {IOP Publishing},
volume = {65},
number = {2},
pages = {026053},
author = {Swanson, C.P.S. and Gates, D.A. and Kumar, S.T.A. and Martin, M.F. and Kruger, T.G. and Dudt, D.W. and Bonofiglo, P.J. and the Thea Energy team},
title = {The scoping, design, and plasma physics optimization of the Eos neutron source stellarator},
journal = {Nuclear Fusion}
}

@article{TJII_Jimenez-Gomez_2011,
doi = {10.1088/0029-5515/51/3/033001},
url = {https://doi.org/10.1088/0029-5515/51/3/033001},
year = {2011},
month = {feb},
publisher = {},
volume = {51},
number = {3},
pages = {033001},
author = {Jiménez-Gómez, R. and Könies, A. and Ascasíbar, E. and Castejón, F. and Estrada, T. and Eliseev, L. G. and Melnikov, A. V. and Jiménez, J.A. and Pretty, D. G. and Jiménez-Rey, D. and Pedrosa, M.A. and de Bustos, A. and Yamamoto, S.},
title = {Alfvén eigenmodes measured in the TJ-II stellarator},
journal = {Nuclear Fusion}
}

@article{Gorelenkov_1999,
    author = {Gorelenkov, N. N and Cheng, C. Z. and Fu, G. Y.},
    title = {Fast particle finite orbit width and Larmor radius effects on low-n toroidicity induced Alfvén eigenmode excitation},
    journal = {Physics of Plasmas},
    volume = {6},
    number = {7},
    pages = {2802-2807},
    year = {1999},
    month = {07},
    issn = {1070-664X},
    doi = {10.1063/1.873545},
    url = {https://doi.org/10.1063/1.873545},
    eprint = {https://pubs.aip.org/aip/pop/article-pdf/6/7/2802/19055052/2802_1_online.pdf},
}

@article{2014Gorelenkov,
  title={Energetic particle physics in fusion research in preparation for burning plasma experiments},
  author={Gorelenkov, N. N. and Pinches, S. D. and Toi, K.},
  journal={Nuclear Fusion},
  volume={54},
  number={12},
  pages={125001},
  year={2014},
  publisher={IOP Publishing}
}

@article{2008Heidbrink,
  title = {Basic Physics of {{Alfv{\'e}n}} Instabilities Driven by Energetic Particles in Toroidally Confined Plasmas},
  author = {Heidbrink, W. W.},
  year = {2008},
  month = may,
  journal = {Physics of Plasmas},
  volume = {15},
  number = {5},
  pages = {055501},
  issn = {1070-664X, 1089-7674},
  doi = {10.1063/1.2838239},
  urldate = {2024-03-25},
  langid = {english}
}

@article{Kolesnichenko2001,
  title = {Alfv{\'e}n Continuum and High-Frequency Eigenmodes in Optimized Stellarators},
  author = {Kolesnichenko, {\relax Ya}. I. and Lutsenko, V. V. and Wobig, H. and Yakovenko, {\relax Yu}. V. and Fesenyuk, O. P.},
  year = {2001},
  month = feb,
  journal = {Physics of Plasmas},
  volume = {8},
  number = {2},
  pages = {491--509},
  issn = {1070-664X, 1089-7674},
  doi = {10.1063/1.1339228},
  urldate = {2024-04-11},
  langid = {english}
}

@book{2021Sharapov,
  title={Energetic particles in Tokamak plasmas},
  author={Sharapov, Sergei},
  year={2021},
  publisher={CRC Press}
}

@article{Weller_1994,
  title = {Neutral beam driven global Alfv\'en eigenmodes in the Wendelstein W7-AS stellarator},
  author = {Weller, A. and Spong, D. A. and Jaenicke, R. and Lazaros, A. and Penningsfeld, F. P. and Sattler, S.},
  journal = {Phys. Rev. Lett.},
  volume = {72},
  issue = {8},
  pages = {1220--1223},
  numpages = {0},
  year = {1994},
  month = {Feb},
  publisher = {American Physical Society},
  doi = {10.1103/PhysRevLett.72.1220},
  url = {https://link.aps.org/doi/10.1103/PhysRevLett.72.1220}
}

\end{document}